\documentclass[12pt,a4paper]{article}
\usepackage[margin=2.54cm]{geometry}
\usepackage[utf8]{inputenc}
\usepackage{amsmath}
\usepackage{graphicx}
\usepackage{subcaption}

\usepackage{amssymb}
\usepackage{amsthm}
\usepackage{xcolor,color}

\usepackage[toc,page]{appendix}

\usepackage{lineno}

\usepackage{mathtools}

\usepackage[breaklinks=true,backref=page]{hyperref}

\usepackage[english]{babel}
\usepackage[square,numbers]{natbib}
\usepackage[nottoc]{tocbibind}

\usepackage{enumitem} 

\usepackage{multirow}

\usepackage{authblk}

\providecommand{\keywords}[1]
{
\small	
\textbf{Keywords:} #1
}

\usepackage{amsthm}
\theoremstyle{plain}
\newtheorem{thm}{Theorem}[subsection]

\theoremstyle{definition}

\theoremstyle{remark}

\title{A Mathematical Model of Helminth Transmission Dynamics under WASH Program Interventions}

\author[1,2]{\small Gonzalo Maximiliano LOPEZ \thanks{gonzalo.maximiliano.lopez@gmail.com}}
\author[1]{\small Juan Pablo APARICIO \thanks{juan.p.aparicio@gmail.com}}
\affil[1]{\small Instituto de Investigaciones en Energía No Convencional (INENCO, UNSa-CONICET),
Universidad Nacional de Salta, Av. Bolivia 5150, A4400FVY, Salta, Argentina}

\affil[2]{\small Departamento de Matemática, Facultad de Cs. Exactas, Universidad Nacional de Salta, Av. Bolivia
5150, A4400FVY, Salta, Argentina}

\date{}
\begin{document}
\maketitle
\begin{abstract}
We present a mathematical model to study the transmission dynamics of soil-transmitted helminth (STH) infections and to assess the impact of community-based water, sanitation, and hygiene (WASH) program interventions. STH infections are a pressing public health issue in vulnerable populations, impairing children’s growth and development. 

Our model explicitly incorporates WASH coverage and effectiveness as dynamic parameters, enabling analysis of their effects on the basic and effective reproduction numbers and the stability of disease-free and endemic equilibria. 

Through saddle-node bifurcation analysis, we identify the critical thresholds in the intervention parameters necessary for infection elimination. 

Numerical simulations show these thresholds and delineate the conditions under which WASH interventions alone may or may not suffice to eliminate transmission, even under widespread coverage. 

Our findings provide a mathematical framework to optimize helminth control strategies and provide evidence-based insights for public health policies aligned with the World Health Organization’s Global Strategy for the elimination of STH infections by 2030.
\end{abstract}
\tableofcontents
\section{Introduction}

Soil-transmitted helminth (STH) infections are significant global public health problems that primarily affect vulnerable populations, particularly children in low-resource communities \cite{hotez2014global}. 
These infections are closely associated with a lack of access to safe drinking water, adequate sanitation, and poor hygiene practices such as inadequate handwashing and direct contact with contaminated soil \cite{freeman2017impact,strunz2014water}. 
As a result, the prevalence and intensity of STH infections are inversely correlated with access to and use of improved water, sanitation, and hygiene (WASH) facilities \cite{freeman2017impact,strunz2014water}. The adverse effects of these infections include anemia, malnutrition, chronic fatigue, and delays in physical and cognitive development, highlighting the need for effective control strategies.

To address this problem, the World Health Organization (WHO) recommends a combined strategy that integrates mass drug administration (MDA) with water, sanitation, and hygiene (WASH) program interventions implemented at the school or community level \citep{paho2003,who2011helminth, who2012soil}.
These initiatives, commonly referred to as school- or community-based WASH programs, aim to reduce exposure to helminth infection and minimize environmental contamination (by eggs and larvae), thereby limiting the spread of infection. 
However, the effective implementation of WASH programs poses significant challenges because their success depends on factors such as population coverage, community adherence to hygiene measures, the effectiveness of interventions, and the availability of infrastructure \cite{clarke2018s,ugwu2024impact}.

Despite these recommendations, evidence regarding the effectiveness of community-based WASH programs remains mixed \citep{ freeman2017impact, garn2022interventions,  nery2019wash,ugwu2024impact,ziegelbauer2012effect}
. 
Some studies conclude that these interventions do not have a significant short-term impact \cite{landeryou2022longitudinal,nery2019wash}, whereas others highlight that high community coverage can reduce transmission \cite{hurlimann2018effect,steinmann2015control}. However, even when benefits are reported, several studies emphasize that community coverage alone does not guarantee a significant impact without the effective implementation of WASH interventions \cite{clasen2014effectiveness,patil2014effect}.

In terms of mathematical modeling, most existing studies evaluate the effect of WASH in combination with MDA without analyzing its independent role in transmission dynamics \cite{clarke2018s, nery2019wash}. 
For example, Coffeng et al. \cite{coffeng2018predicted} modeled community-based WASH interventions alongside ongoing deworming efforts. They concluded that although interventions have limited short-term impact, they are essential to prevent infection recurrence after deworming suspension.

Given the lack of mathematical models that exclusively evaluate the impact of community-based WASH, we develop a deterministic compartmental model based on differential equations that explicitly incorporates the coverage and effectiveness of these interventions as key parameters. 

Stability and bifurcation analyses are used to identify critical thresholds for infection elimination and to examine the impact of heterogeneity in adherence to interventions on transmission. 
The study focuses on the equilibria of the system and their stability and identifies the coverage and effectiveness thresholds necessary to achieve an infection-free equilibrium. Through a bifurcation analysis, we demonstrate the existence of a saddle-node bifurcation and determine the impact of these thresholds on transmission dynamics. 

By analyzing the basic reproduction number (\( R_0 \)) and the effective reproductive number (\( R_e \)), we establish quantitative criteria to evaluate the feasibility of eradication \citep{churcher2006density, diekmann2000mathematical}.
Furthermore, bifurcation diagrams are used to identify the conditions under which the infection persists in the population, with direct implications for intervention strategy design.

We also perform numerical simulations to explore various intervention scenarios and assess their potential to eliminate infection.

In summary, this work provides a theoretical framework to evaluate the standalone impact of community-based WASH programs on helminth transmission. 
Our results offer insights into the critical thresholds required for elimination, guide the development of more effective control strategies, and support ongoing global efforts to achieve the WHO goal of eliminating helminthiasis as a public health problem by 2030 \cite{who2019sth2030,who2021roadmap}.

\section{Mathematical Model for the Transmission Dynamics of Helminth Infections}  
\label{s:model1} 

\subsection{Model Formulation}
A basic model for the transmission dynamics of helminth infections was introduced by Anderson and May in their seminal 1985 paper \citep{anderson1985helminth,anderson1992infectious}. 
This model focuses on the interaction between two key variables: the \textbf{mean parasite burden} in the host population ($M$), which represents the mean number of parasites per infected host, and the \textbf{infective environment} ($L$), which captures the concentration of infective stages (such as eggs or larvae) in the external environment. These variables are coupled by a set of nonlinear ordinary differential equations (ODEs) that describe the dynamic relationship between the host-parasite system and the external environment
\begin{equation}\label{model1}
\begin{split}
\dfrac{dM}{dt}&=\beta L - (\mu_H+\mu_P) M,\\
\dfrac{dL}{dt}&= \alpha \lambda_0  \rho M F(M)  - \mu_L L.
\end{split}
\end{equation}
The associated parameters are defined as follows:
\begin{itemize}
\item $\beta$ the contact rate (or exposure rate), which quantifies how often hosts come into contact with infective stages in the environment.
\item $\rho$ is the contribution rate of a host to reservoir $L$, representing the rate at which hosts contribute infectious stages to the environment.
\item $\mu_H$, $\mu_P$ and $\mu_L$ are the mortality rates associated with the host, the parasite, and the environmental reservoir, respectively.
\item $\alpha$ represents the proportion of females in the parasite population (sex ratio).
\item $\lambda_0$ the egg production rate per female parasite, independent of parasite density within the host.
\item $F(M)$ is a product of functions that quantify density-dependent processes affecting parasite fecundity and mating probability.
\end{itemize}

The function $F(M)$ is a key component that captures the nonlinear effects of parasite density on fecundity and mating dynamics within the host for a given parasite distribution \cite{lopez2024modeling}. It is the product of two functions:

\begin{itemize}
    \item $\psi(M)$: quantifies the density-dependent fecundity of parasites, accounting for the distribution of parasites in the host population \cite{lopez2024modeling}.
    \item $\phi(M)$: describes the mating probability of parasites, assuming a polygamous mating system and counting for the distribution of parasites within the host \cite{lopez2024modeling}.
\end{itemize}

When a \textbf{negative binomial distribution} for the parasite burden within the host population is assumed, the specific forms of $\psi$ and $\phi$ are given as follows \cite{lopez2024modeling,lopez2022modeling}:
\begin{align}
\label{psi}\psi(M;z,k)&=\left[ 1+(1-z) \tfrac{M}{k}\right] ^{-(k+1)},\\
\label{phi}\phi(M;z,k)&=1-\left[ \dfrac{1+(1-\alpha z)\frac{M}{k}}{1+(1-z)\frac{M}{k}}\right] ^{-(k+1)},
\end{align}
where $M$ is the mean parasite burden and $k$ is the inverse of the parasite dispersion parameter, both parameters correspond to a negative binomial model, and $z = e^{-\eta}$ models the density-dependent fertility rate decline, where $\eta$ represents the intensity of this decrease. The density-dependent fecundity is described by $\lambda_0 z^{n-1}$,where $n$ is the number of parasites in the host \cite{hall2000geographical}.

\subsection{Reproduction Numbers and Model Dynamics}
\label{ss:dinamic_model1}

\begin{figure}
\centering
\includegraphics[width=0.5\linewidth]{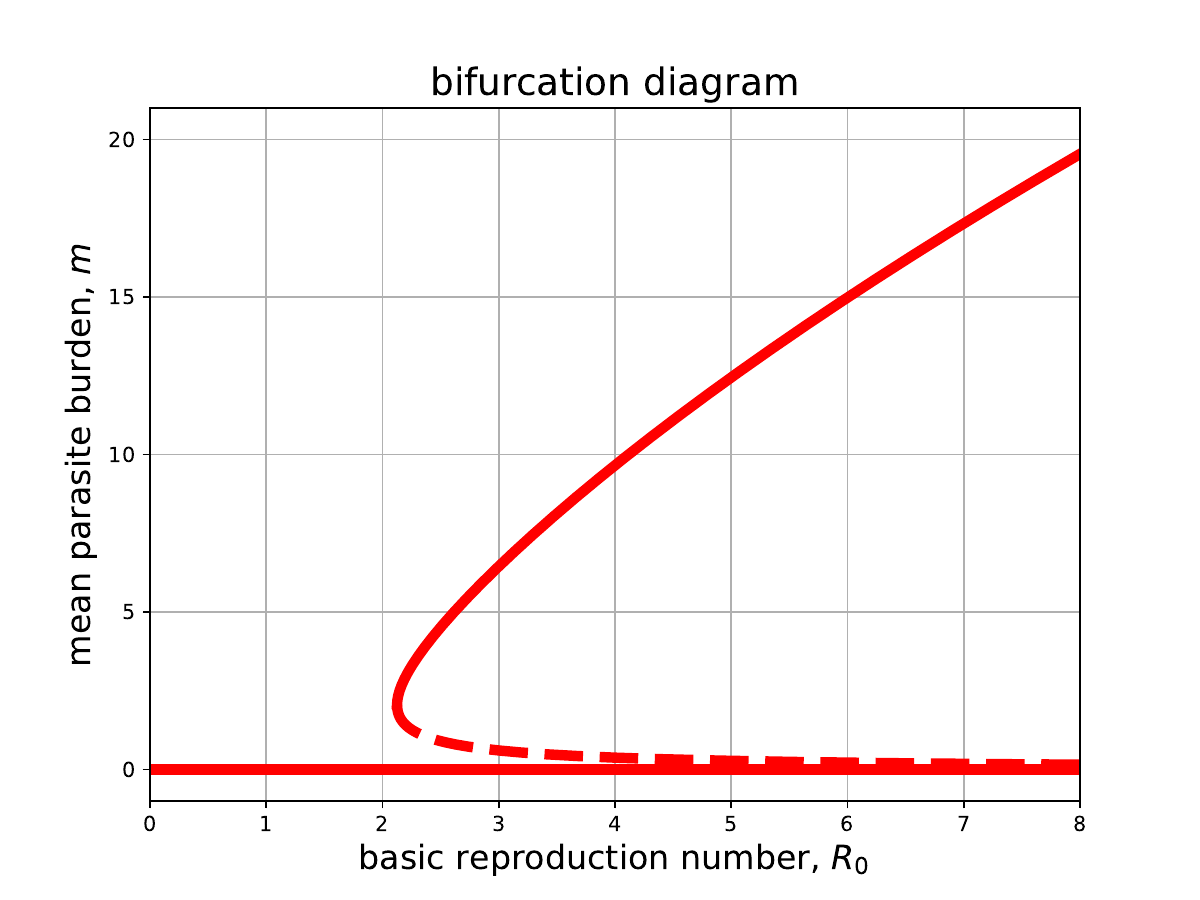}
\caption{
Bifurcation diagram illustrating the transition between disease-free and endemic equilibria as the basic reproduction number $R_0$ varies. The vertical axis represents the mean parasite burden, $M$. As $R_0$ increases past a critical value of $R_0^b$, saddle-node bifurcation occurs, leading to the emergence of two non-trivial equilibria: one stable and one unstable. The parameter values used where $\alpha = 0.574$, $k = 0.7$, and $z = 0.92$ \citep{anderson1992infectious,seo1979egg}.
}
\label{f:plot_bifur}
\end{figure}

The model assumes that the duration of other stages in the parasite's life cycle is short compared to the life expectancy of an adult parasite \cite{anderson1985helminth,anderson1992infectious}. It also assumes that no mortality occurs during these earlier stages. 

Moreover, reservoir dynamics is assumed to be much faster than parasite-host dynamics. 
Based on the adiabatic elimination of the fast variables, this simplification allows us to assume that the variable \( L \) is always at equilibrium \cite{strogatz2024nonlinear}. 
Therefore, the two-dimensional system \eqref{model1} reduces to the one-dimensional system:
\begin{equation}
\label{model1a}
\frac{dM}{dt} = (\mu_H + \mu_P) \left[ R_0 F(M) - 1 \right] M,
\end{equation}
where $M$ denotes the mean parasite burden per host, $\mu_H$ and $\mu_P$ are the host and parasite mortality rates, respectively, \( R_0 \) is the basic reproduction number, and $F(M) = \psi(M)\phi(M)$ captures the density-dependent fecundity and mating success.

In this equation, $1/(\mu_H + \mu_P )$ represents the life expectancy of the adult parasite
in the human host, taking into account the combined parasite and human mortality.

The effective reproduction number, denoted as \(R_e\), is defined as \(R_e = R_0 F(M)\). 
model~\eqref{model1a} clearly shows that the threshold for macroparasite spread occurs when \(R_e = 1\). 
When \(R_e < 1\), the infection cannot persist in the host population and eventually disappears.  
When \(R_e > 1\), the infection is sustained in the host population and reaches its endemic state. 

However, it is important to highlight that \(R_0 > 1\) alone is not sufficient to ensure the existence of an endemic equilibrium, due to mating limitations at low parasite burdens (see Figure \ref{f:plot_bifur})\cite{lopez2024modeling}. 

The dynamical system  that describes helminth infection dynamics with density-dependent effects is well studied and exhibits a \textbf{saddle-node bifurcation} at  a critical point denoted by \((M^b, R_0^b)\) \cite{anderson1992infectious,lopez2022modeling,lopez2024modeling}.

When male and female parasites are distributed together within the same host, the following equations define the conditions for such a bifurcation \cite{lopez2024modeling}. 
These equations also allow for the determination of the bifurcation values \(M^b\) and \(R_0^b\):  
\begin{equation}
f(M, R_0) = 0, \quad Df(M, R_0) = 0, \quad D^2f(M, R_0) \neq 0, \quad f_{R_0}(M, R_0) \neq 0,
\end{equation}
where \(f\) represents the right-hand side of the dynamical system \eqref{model1a}, $\frac{dM}{dt} = f(M, R_0)$.
According to Sotomayor's Theorem, these conditions ensure the existence of a saddle-node bifurcation in the dynamical model \cite{chicone2006ordinary,perko2013differential,sotomayor1973generic}.

Analysis of the deterministic model \eqref{model1a} reveals two key scenarios:

\begin{itemize}

\item If \(R_0 < R_0^b\) (\(R_e < 1\)), the system has a single stable equilibrium: the disease-free equilibrium at \(M = 0\).

\item When \(R_0 > R_0^b\) (\(R_e > 1\)), two non-trivial equilibria emerge (see Figure \ref{f:plot_bifur}). 

The first equilibrium is unstable and acts as a threshold that separates stable \textbf{endemic equilibrium} (representing persistent infection) from stable \textbf{disease-free equilibrium} (representing parasite extinction). This unstable equilibrium is commonly referred to as the \textbf{transmission breakpoint}. 
The term highlights the dioecious nature of the parasites, which require both male and female parasites to be present in the same host to produce viable eggs and maintain the environmental reservoir of infectious material.

The second equilibrium occurs when \(R_0 > R_0^b\) and \(M\) exceeds the breakpoint. 
This equilibrium, referred to as the stable endemic equilibrium, represents the persistence of infection (see Figure \ref{f:plot_bifur}).

\end{itemize}

\section{Extended Model Incorporating Community-Based WASH Program Interventions}
\label{s:model2}

\subsection{Introduction}
The World Health Organization (WHO) recommends implementing WASH (water, sanitation, and hygiene) programs as a complementary strategy to mass drug administration (MDA) for controlling parasitic infections \citep{who2012soil,who2011helminth}.
Improved access to safe drinking water, such as protected wells, reduces exposure to pathogens. Likewise, proper sanitation infrastructure, including ventilated latrines and septic tanks, ensures the safe disposal of human waste, disrupting transmission pathways.
Promoting hygiene practices, such as regular handwashing and personal cleanliness, is also essential for preventing disease. In endemic areas, wearing footwear protects against contact with contaminated soil and helps prevent parasitic infections \citep{paho2003, who2012soil}.

\subsection{Improved Sanitation}

Ensuring access to safe water and adequate sanitation is a critical public health measure to reduce soil contamination with infective helminth eggs. 
These interventions include the construction of ventilated latrines, septic systems, and wastewater treatment facilities to prevent human waste from entering the environment. By reducing the presence of helminth eggs and larvae in soil and water, these measures disrupt the parasite transmission cycle and lower the incidence of infection in community hosts.

However, these improvements often require significant structural changes, including proper infrastructure and substantial financial resources, which may not be feasible in low-resource settings. In low- and middle-income countries, a lack of funding, weak infrastructure, and limited investment in sanitation services have hindered the success of these interventions.

Sanitation improvements have successfully interrupted helminth and other fecal-oral transmitted diseases in high-income countries \citep{nery2019wash}, but their impact in low- and middle-income countries has been limited \citep{garn2022interventions}. This underscores the need for targeted, sustained implementation efforts adapted to local conditions.

\subsection{Improved Hygiene}

Promoting hygiene practices in conjunction with access to safe water is a crucial strategy to reduce helminth transmission and reinfection \citep{puspita2020health}. Efforts include encouraging handwashing with soap,  especially after defecation and before eating, and wearing shoes in endemic regions to prevent hookworm transmission (e.g., Ancylostoma and Necator species). These behaviors help prevent direct contact with helminth eggs and larvae present in contaminated environments.

Personal hygiene also involves handwashing, keeping nails clean and short, and avoiding direct contact with potentially contaminated soil or water. Health education campaigns promote these habits by providing people with the knowledge and tools necessary to protect themselves from parasitic infections.

The success of these interventions depends heavily on community participation and continuous education about hygiene practices. 
Although educational strategies are often less expensive than infrastructure projects, their effectiveness is based on achieving lasting behavioral change. 
This is particularly relevant in low- and middle-income countries, where socioeconomic and cultural factors can hinder the adoption of healthy hygiene habits.

\subsection{Community-Based WASH Program Concepts in the Model}

For modeling purposes, it is important to consider that WASH program interventions do not act in isolation. Instead, they complement each other to maximize their impact on reducing helminth transmission. We categorize WASH programs into three modalities:

\begin{itemize}
\item \textbf{Sanitation modality}: This includes interventions such as access to safe water and adequate sanitation facilities. These  interventions reduce the host's contribution to the environmental reservoir (e.g., through latrine use and wastewater management).

\item \textbf{Hygiene modality}: This includes interventions such as access to safe water and hygiene education. These reduce the host’s exposure to infection (e.g., handwashing, personal hygiene, and wearing shoes).

\item \textbf{WASH modality}: This represents the combined impact of sanitation and hygiene interventions.
\end{itemize}

The impact of these interventions is quantified using two parameters: \textbf{coverage} and \textbf{effectiveness}.

Here, \textbf{coverage} refers to the proportion of the host population that has been reached and agreed to participate in WASH programs.

\textbf{Effectiveness}, on the other hand, refers to the average reduction in exposure or contribution to transmission over time, assuming partial or full adoption of WASH interventions by individuals. However, this parameter does not capture potential limitations such as irregular or improper use.



\subsection{Model Formulation}

To evaluate the impact of water, sanitation, and hygiene (WASH) programs on a host community, we analyzed their effects on the parasite burden within the host community. 
Building on a previous model that considered parasite transmission in a homogeneous host community, we extended it to incorporate the effects of WASH program interventions.

In this extended model, the host community is divided into two groups: individuals who benefit from WASH program interventions, denoted by $H_1$, and those who do not, denoted by $H_0$. 

This division reflects a scenario in which a proportion $c$ of the population is affected by the intervention, whereas the remaining proportion $1 - c$ remains unaffected. The parameter $c$ represents the \textbf{coverage} level of the intervention. The total host community, $N$, is given by $N = cN + (1 - c)N$.

The dynamics of the model, incorporating WASH program interventions, are described by the following equations:
\begin{equation}
\label{model2}
\begin{split}
\dfrac{dM_k}{dt}&=\beta_k L - (\mu_H+\mu_P) M_k\\
\dfrac{dL}{dt}&= \alpha \lambda_0 \left[ \sum_k \rho_k q_k M_k F(M_k)\right]   - \mu_L L ,
\end{split}
\end{equation}
where $k=0,1$.

Here, $M_1$ and $M_0$ represent the mean parasite burden in individuals with and without access to WASH programs, respectively.

The impact of WASH programs is incorporated through two key parameters: \textbf{coverage} and \textbf{effectiveness}. 
The coverage parameter $c$ defines the weight assigned to each subpopulation via
\begin{equation}
q_k = c k + (1 - c)(1 - k) \quad\text{where $k=0,1$}.
\end{equation}
The effectiveness of sanitation ($e_s$) and hygiene ($e_h$) interventions is reflected in the modified transmission parameters:
\begin{equation}
\beta_k = \beta(1 - e_h)k + \beta(1 - k), \quad 
\rho_k = \rho(1 - e_s)k + \rho(1 - k)
\quad\text{where $k=0,1$}.
\end{equation}
Here, \(e_h\) and \(e_s\) represent the reduction in transmission due to hygiene and sanitation interventions, respectively.


\section{Local Dynamics of Extended Model}

In this section, we explore the equilibrium states and local dynamics of a system that models the helminth transmission under the influence of WASH program interventions. 
Since  the WASH programs affect the environmental and behavioral  transmission components, which are modeled through the parameters \(\beta\) (contact rate) and \(\rho\) (contribution rate), we assume that these parameters remain constant after implementation, leading to a modified basic reproduction number \(\hat R_0\).

The analysis focuses on two key epidemiological thresholds: the \textbf{basic reproduction number} (\(R_0\)) and the \textbf{effective reproduction number} (\(R_e\)). While \(R_0\) provides a theoretical threshold for parasite invasion in a fully susceptible population without density-dependent constraints, \(R_e\) accounts for dynamic interactions such as mating probabilities and parasite crowding, which generate a nonlinear relationship with parasite burden.

We examine the conditions under which disease-free, endemic, or unstable equilibria may arise by incorporating population heterogeneity and density dependence. Then, we derive expressions for these equilibria within our model and characterize the bifurcation point that determines the critical threshold for parasite persistence or elimination.

Through this analysis, we aim to gain deeper insights into how WASH program interventions modulate parasite transmission dynamics and the critical conditions that govern the long-term behavior of the infection within a host community.

\subsection{Equilibria and Reproduction Numbers}
\subsubsection{Basic Reproduction Number}
The basic reproduction number (\(R_0\)) is a fundamental concept in parasitology. For microparasites, \(R_0\) is defined as the average number of secondary infections generated by a primary infection in the absence of density-dependent constraints \cite{anderson1992infectious, diekmann2000mathematical, diekmann1990definition}. It serves as a threshold parameter: an epidemic occurs only if \(R_0 > 1\).

For macroparasites, \(R_0\) represents the average number of female offspring produced by a single mature female during her reproductive life, assuming no density-dependent factors \cite{anderson1992infectious}. 

Understanding \(R_0\) is essential because it serves as an idealized threshold for parasite invasion and persistence within a host community. Unlike with microparasites, where an \(R_0 = 1\) serves as a clear threshold for epidemic outbreaks, for macroparasites \(R_0 = 1\) does not represent such a strict threshold due to the complex interplay of density-dependent factors discussed in Section~\ref{s:model1}. This metric is critical for understanding macroparasite transmission dynamics and designing effective control strategies.\cite{anderson1992infectious}.

\subsubsection{Effective Reproduction Number}

The effective reproduction number (\(R_e\)) is another critical factor in the macroparasite transmission dynamics. It accounts for parasite density and density-dependent effects, resulting in a characteristic \textbf{hump-shaped} relationship when plotted against parasite burden \cite{anderson1992infectious,basanez2012research,churcher2006density}. Understanding \(R_e\) is crucial because it provides insight into parasite transmission dynamics and population persistence.

This behavior is driven by \textbf{positive density-dependence} (facilitation), also known as the \textbf{Allee effect}, where an increasing parasite burden facilitates transmission by increasing the likelihood that female parasites are mated. This is followed by \textbf{negative density-dependence} (limitations), where high parasite burdens limit per capita rates of parasite establishment, survival, and fecundity \cite{churcher2006density,lopez2024modeling}.

These constraints cause \(R_e\) to reach a value of 1 at two different parasite densities. 
The density to the right side of the \textbf{hump} corresponds to the \textbf{endemic equilibrium}, where each parasite, on average, replaces itself, maintaining the population through negative density-dependence (limitations). 
In contrast, the density on the left side of the \textbf{hump} represents an \textbf{unstable equilibrium}, known as the \textbf{transmission breakpoint} density, governed by positive density-dependence (facilitation). 
At this transmission breakpoint density, the parasite population cannot sustain itself and will eventually  become  extinct, although this process may be slow. These points, where \(R_e = 1\), define critical thresholds for population persistence or collapse (see Figure \ref{f:plot_bifur}).

\subsubsection{Basic Reproduction Number in the Extended Model}

In this section, we describe the basic reproduction number in the context of our model, which includes the WASH program interventions. 
Before the implementation of this program, the basic reproduction number for the classical Anderson-May model of macroparasite transmission dynamics is given by:
\begin{equation}
R_0 = \frac{\alpha \lambda_0  \beta \rho}{\mu_L(\mu_H + \mu_P)}.
\end{equation}
This corresponds to the basic reproduction number of the initial model \ref{model1}, as presented in Section \ref{s:model1} and discussed in previous works \cite{anderson1992infectious, lopez2022modeling, lopez2024modeling}.

To compute the basic reproduction number for our model with WASH program interventions, we assume that these interventions are maintained over time once initiated, which leads to changes in the parasite transmission dynamics within the host community.

The implementation of the WASH program transforms a previously homogeneous population into a heterogeneous population, consisting of treated and untreated individuals.

Although the entire population remains susceptible to infection, transmission conditions differ between  these two host groups (treated and untreated).
Based on this assumption, we introduce a new basic reproduction number, denoted as $\hat{R}_0$.

As mentioned by \cite{chan1994development, heesterbeek1995threshold}, in the context of parasite transmission dynamics in a heterogeneous host population, the basic reproduction number for each host group, characterized by a mean parasite burden $M_i$, is given by:
\begin{equation}
\hat R_{0,i} = \frac{\alpha \lambda_0 \beta_i \rho_i q_i}{\mu_L(\mu_H + \mu_P)}
\quad\text{where $k=0,1$}.
\end{equation}
The basic reproduction number for the new dynamical system with the WASH programs, denoted as \(\hat R_0\), is then given by:
\begin{equation}
\label{hatR0}
\hat R_0 = \frac{\alpha \lambda_0}{\mu_L (\mu_H + \mu_P)} \sum_i \beta_i \rho_i q_i .
\end{equation}
This expression can be interpreted as the basic reproduction number of the original system \ref{model1}, scaled by the factor \(\frac{\sum_i \beta_i \rho_i q_i}{\beta \rho}\).

To evaluate the impact of the WASH program interventions, we assume that, in the host group where the program is applied, the initial contact rate $\beta$ is reduced by a proportion $e_h$ (hygiene effectiveness), and the contribution rate $\rho$ is reduced by a proportion $e_s$ (sanitation effectiveness). Based on these assumptions, the basic reproduction number for the system \ref{model2} can be expressed as follows, according to equation \eqref{hatR0}:
\begin{equation}
\hat R_0 = R_0 \left[ 1 - c(e_h + e_s - e_h e_s) \right],
\end{equation}
where \(R_0 = \frac{\alpha \lambda_0 \beta \rho}{\mu_L(\mu_H + \mu_P)}\) corresponds to the basic reproduction number of the initial parasite transmission dynamic system \ref{model1}. The relation between \(\hat R_0\) and \(\hat R_{0,i}\) is thus
\begin{equation}
\hat R_0 = \sum_i \hat R_{0,i}
\end{equation}


\subsubsection{Equilibria}
\label{ss:equilibria}

Similar to the approach described in Section~\ref{ss:dinamic_model1}, we simplify model~\eqref{model2} to determine the equilibria of the system dynamics influenced by WASH programs.

Since that the lifespan of adult parasites in human hosts is significantly longer than that of larvae or eggs, the rate of change of \(L\) is expected to be much faster compared to the parasite loads \(M_0\) and \(M_1\). Therefore, under the assumption that \(L\) has reached its equilibrium state, we rewrite the model~\eqref{model2} as follows (see section~\ref{ss:dinamic_model1}):    
\begin{equation}
\label{model2a}
\dfrac{dM_i}{dt} = (\mu_H+\mu_P)
\left[ \frac{ R_0 \beta_i }{\beta \rho } 
\left( \sum_j \rho_{j} q_{j} M_{j} F(M_j)\right) - M_i \right], 
\end{equation}
where \(i, j = 0, 1\).

The following Theorem characterizes the non-trivial equilibra of the model~\eqref{model2a}, which corresponds to states where the parasite populations are non-zero.

\begin{thm}
\label{thm:equilibria}
The non-trivial equilibria \(E=(M_0^*,M_1^*)\) of model~\eqref{model2a} (i.e., \(M_i^* \neq 0\)) are characterized by the following conditions:
\begin{equation}
\sum_i \hat R_{0,i} F(M_i)=1
\qquad \text{and} \qquad 
\frac{M_{0}}{\beta_{0}} = \frac{M_{1}}{\beta_{1}},
\end{equation}
where \(i=0,1\), and \(R_0\) is the basic reproduction number of the initial model~\eqref{model1}.
\end{thm}



\subsubsection{Effective Reproduction Number in the Extended Model}
\label{ss:Re}

Following the classical Anderson and May framework \citep{anderson1992infectious,anderson1985helminth}, we define the effective reproduction number in order to analyze the system's equilibria and gain a better understanding of its dynamics (see also Section \ref{s:model1}).

We begin by defining the auxiliary function
\begin{equation}
g(m_0, m_1) = \frac{\sum_i \beta_i \rho_{i} q_{i}}{\beta \rho} F(m_i) , 
\end{equation}
from which we derive the function 
\( G(m) = g\left(m, \frac{\beta_1 m}{\beta_0}\right) \). 
The graph of $G(m)$ as a function of $m$ has a characteristic hump-shaped form \cite{basanez2012research,churcher2006density}.
Non-trivial equilibria exist where this curve, scaled by $R_0$, intersects the constant horizontal line at unit height (i.e., $y = 1$).

The \textbf{effective reproduction number} is defined as 
\begin{equation}
R_e = R_0 G(m),
\end{equation}
incorporating the non-linear effects of parasite density.

For non-trivial equilibria to exist, the hump of this function must intersect the constant horizontal line at unit height, which implies the following condition:
\begin{equation}
R_e = R_0 G_{\max} \geq 1,
\end{equation}
where \( G_{\max} = \max_{m \geq 0} G(m) \).

As in the classic Anderson and May models, this condition is necessary for the infection to persist (see Section~\ref{s:model1}). Solving for $R_0$, we obtain the threshold value at which non-trivial equilibria emerge:
\begin{equation}
R_0 G_{\max} = 1.
\end{equation}
Thus, the threshold value of $R_0$ for the existence of non-trivial equilibria is given by: 
\begin{equation}\label{R0b}
R_0^b = \frac{1}{G_{\max}}=\frac{1}{G(m^b)},
\end{equation}  
where $m^b$ denotes the value of $m$ at which the function $G(m)$ attains its maximum. 

In our model, the bifurcation point is defined as $E^b = (M_0^b, M_1^b)$, where $M_0^b = m^b$ and $M_1^b = \frac{\beta_1 m^b}{\beta_0}$. This condition is consistent with the non-trivial equilibria of our system (see Theorem~\ref{thm:equilibria}).

In the absence of WASH program interventions, the model recovers the bifurcation point described in the basic model of Section~\ref{s:model1}, as referenced in previous studies \cite{anderson1992infectious,lopez2022modeling,lopez2024modeling}.

\subsection{Disease-Free and Endemic Equilibria}

The analysis of equilibrium provides crucial insights into the long-term behavior of the infection dynamics. In particular, we investigate the conditions under which the parasite infection is eradicated from the host community (i.e., disease-free equilibrium) and when it persists at a stable, non-zero level (i.e., endemic equilibrium). 

The existence and nature of these equilibria are governed by the basic reproduction number $R_0$, and more specifically, by its comparison with the critical threshold $R_0^b$, which emerges from the nonlinear structure of the model and the properties of the function $G(m)$.

The following Theorem characterizes the number and type of equilibria of system \eqref{model2a} depending on the value of $R_0$ , highlighting the occurrence of a bifurcation when $R_0 = R_0^b$.

\begin{thm}
\leavevmode
\begin{enumerate}[label=(\roman*)] 
\item If \( R_0 < R_0^b \), the model system \eqref{model2a} admits a unique equilibrium: the \textbf{disease-free equilibrium} \(E^0 = (0, 0)\).

\item If \( R_0 > R_0^b \), the model system \eqref{model2a} admits three equilibria: disease-free equilibrium \( E^0 \),  stable \textbf{endemic equilibrium} \( E^s \), and unstable equilibrium \( E^u \), known as the \textbf{transmission breakpoint}.

\item If \( R_0 = R_0^b \), the model system \eqref{model2a} admits two equilibria: the disease-free equilibrium \( E^0 \) and the \textbf{bifurcation equilibrium} \( E^b \).
\end{enumerate}
\end{thm}

\subsection{Local Stability Analysis}

To better understand the qualitative behavior of the system near the equilibrium, we perform a local stability analysis. This involves evaluating the Jacobian matrix at each equilibrium and analyzing the sign of its eigenvalues. Depending on the value of the basic reproduction number \( R_0 \), the stability properties of the disease-free and endemic equilibria vary, revealing important thresholds for parasite persistence or elimination.

\begin{thm}
\leavevmode
\begin{enumerate}[label=(\roman*)] 
\item If \( R_0 < R_0^b \), the \textbf{disease-free equilibrium} \( E^0 = (0, 0) \) is locally asymptotically stable.
\item If \( R_0 > R_0^b \), the \textbf{disease-free equilibrium} \( E^0 \) remains locally asymptotically stable, the \textbf{endemic equilibrium} \( E^s \) is locally asymptotically stable, and the equilibrium \( E^u  \) (\textbf{transmission breakpoint}) is unstable.
\end{enumerate}
Here, \( R_0^b \) was introduced in Section~\ref{ss:Re} and defined in Equation~\eqref{R0b}.
\end{thm}

\subsection{Saddle-node Bifurcation}

This section examines a \textbf{saddle-node bifurcation} in system~\eqref{model2a}, which plays a critical role in understanding the emergence and disappearance of endemic equilibria. In particular, we focus on how changes in the basic reproduction number \( R_0 \) affect the number and stability of equilibria. A saddle-node bifurcation occurs when two equilibria, typically one stable and one unstable, coalesce and annihilate each other as a parameter (here, \( R_0 \)) crosses a critical threshold.

We apply \textbf{Sotomayor's Theorem} \cite{sotomayor1973generic}, to rigorously establish the existence of this bifurcation near the critical value \( R_0 = R_0^b \), where the endemic and transmission-breakpoint equilibria merge.

\begin{thm}
    The model~\eqref{model2a} undergoes a saddle-node bifurcation around $E^b$ at $R_0=R_0^b$.
\end{thm}

\section{Impact of WASH Program Parameters on Helminth Transmission Dynamics}
\label{s:impact}
\subsection{Effective Reproduction Number and WASH Program Parameters}
\begin{figure}[h!]
\centering
\includegraphics[width=0.7\linewidth]{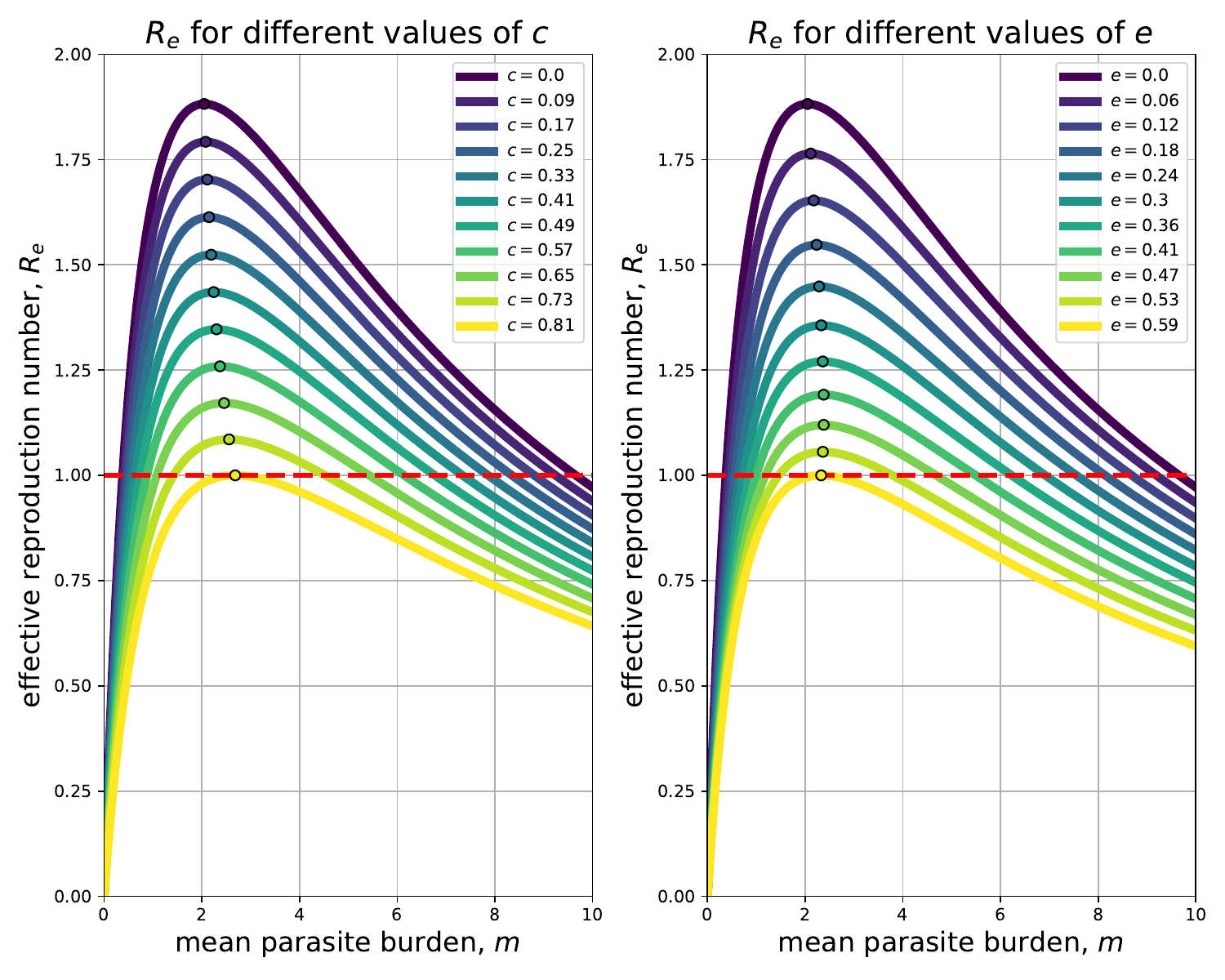}
\caption{
The graph on the left shows the variation of the effective reproduction number ($R_e$) as a function of the population coverage of WASH programs ($c$), assuming fixed values for the basic reproduction number ($R_0$), sanitation effectiveness ($e_s$), and hygiene effectiveness ($e_h$). The intersections with the horizontal line at a value of one ($R_e = 1$) represent system equilibria. The critical point indicates a saddle-node bifurcation that occurs at the minimum coverage threshold ($c_{\min}$) required to eliminate infections. The graph on the right extends the analysis to include variations in $e_s$ and $e_h$, assuming $e_s = e_h = e$, to evaluate integrated intervention strategies. Parameters used: $R_0 = 4$; for the left graph, $e = 0.35$; for the right graph, $c = 0.55$.
}
\label{fig:Re_vs_c_and_e}
\end{figure}

In this section, we analyze the impact of  WASH program parameters, such as the population coverage ($c$) and the effectiveness of sanitation ($e_s$) and hygiene ($e_h$) interventions, on the variation of the Effective Reproduction Number ($R_e$). 

This analysis enables us to evaluate the effects of WASH programs on the model dynamics and to determine the optimal conditions for achieving a disease-free state in the system.

The Effective Reproduction Number ($R_e$) is related to the Basic Reproduction Number ($R_0$) by the following equation:
\begin{equation}
	R_e = R_0 G(m),
\end{equation}
Here, G(m) describes a characteristic hump-shaped curve reflecting nonlinear interactions in parasite transmission
\cite{anderson1992infectious,basanez2012research,churcher2006density,lopez2024modeling}. 

The parameters $c$, $e_s$, and $e_h$ explicitly affect $G(m)$ through the following relation:
\begin{equation}
	G(m;c,e_s,e_h) = (1-c)F(m) + c(1-e_s)(1-e_h)F((1-e_h)m),
\end{equation}
where $F(m)$ is related to parasite density in the population, as defined in Section~\ref{ss:Re}.

Since $G(m;c,e_s,e_h)$ depends on $c$, $e_s$, and $e_h$, both $m^b(c,e_s,e_h)$, the value of $m$ at which $G(m;c,e_s,e_h)$ reaches its maximum, and the saddle-node bifurcation threshold $R_0^b(c,e_s,e_h)$ also depends on these parameters. This threshold is given by:
\begin{equation}  
	R_0^b = \frac{1}{G(m^b;c,e_s,e_h)},  
\end{equation}  
where $m^b$ satisfies the condition:
\begin{equation}  
	G'(m^b;c,e_s,e_h) = 0.  
\end{equation}

The equilibria of the system are determined by the following  equation:
\begin{equation}
	R_0 G(m;c,e_s,e_h) = 1,
\end{equation}  
This corresponds to the points at which the curve $R_0 G(m;c,e_s,e_h)$ intersects the constant horizontal line at unit
height.

Note that the value of $R_0$ determines the height of the curve $R_0 G(m;c,e_s,e_h)$ (see Figure~\ref{fig:Re_vs_c_and_e}).  
When $R_0 = R_0^b(c,e_s,e_h)$, the system undergoes a saddle-node bifurcation in which two equilibrium solutions, one stable and one unstable, collapse (i.e., the curve $R_0 G(m;c,e_s,e_h)$ intersects the unit-height constant horizontal line at a single point).  

Given a fixed value of $R_0$ and assuming constant values for $e_s$ and $e_h$, we can determine a critical value $c_{\min}$ such that, for $c \in [0, c_{min}]$, the system exhibits multiple equilibria. The value $c_{\min}$ corresponds to the \textbf{minimum WASH coverage} required to push the system to the \textbf{saddle-node bifurcation threshold}, where the stable and unstable endemic equilibria collapse. Although we refer to it as $c_{\min}$ because it is the upper limit of the interval considered, from a control perspective, it represents the \textbf{ minimum coverage needed} to eliminate the infectious disease. 
Beyond this point, only the disease-free equilibrium remains.

With $R_0$, $e_s$, and $e_h$ fixed, the system equilibria can be plotted as a function of $c$ over the interval $[0, c_{\min}]$:  
\begin{itemize}
	\item $c = 0$ : Absence of WASH programs ( i.e., no interventions).  
	\item $c = c_{\min}$ : The critical point at which the saddle-node bifurcation occurs, and $R_0^b(c,e_s,e_h)$ equals the fixed value of $R_0$.
\end{itemize}
These conditions are described mathematically as follows:
\begin{equation}
	\begin{split}
		R_0 G(m;c) &= 1, \\
		G'(m;c) &= 0.
	\end{split}
\end{equation}  

Figure~\ref{fig:Re_vs_c_and_e} shows the variation of the Effective Reproduction Number $R_e = R_0G(m;c)$ with respect to $c$. The points at which the curve intersects the unit-height constant horizontal line represent the system equilibria. This illustrates how WASH programs modify the model dynamics through the parameter c.
 
Similarly, the analysis can be extended by considering variations in $e_s$ and $e_h$. This provides a comprehensive framework to assess the effectiveness of WASH interventions in transmission dynamics and strategies to achieve a disease-free state. 
For simplicity, we will consider a single value for the parameters $e_s$ and $e_h$, denoted by $e$. Figure~\ref{fig:Re_vs_c_and_e} illustrates the behavior of $R_e$ as a function of this parameter.


\subsection{Parametric Analysis of WASH Program Impact}

As previously discussed, the impact of WASH program interventions on helminth infection transmission dynamics can be assessed through the effective reproduction number ($R_e$). This quantity is influenced by three key parameters: the intervention coverage ($c$), the intervention effectiveness ($e$), and the baseline transmission potential represented by the basic reproduction number ($R_0$). In what follows, we perform a parametric analysis to evaluate how variations in each of these factors affect the equilibrium outcomes of the system and the feasibility of infection elimination. Finally, we consider the role of biological heterogeneity by incorporating the parasite aggregation parameter ($k$) in Section \ref{sss:k}.

\subsubsection{Coverage} 
\begin{figure}[h!]  
\centering  
\includegraphics[width=0.7\linewidth]{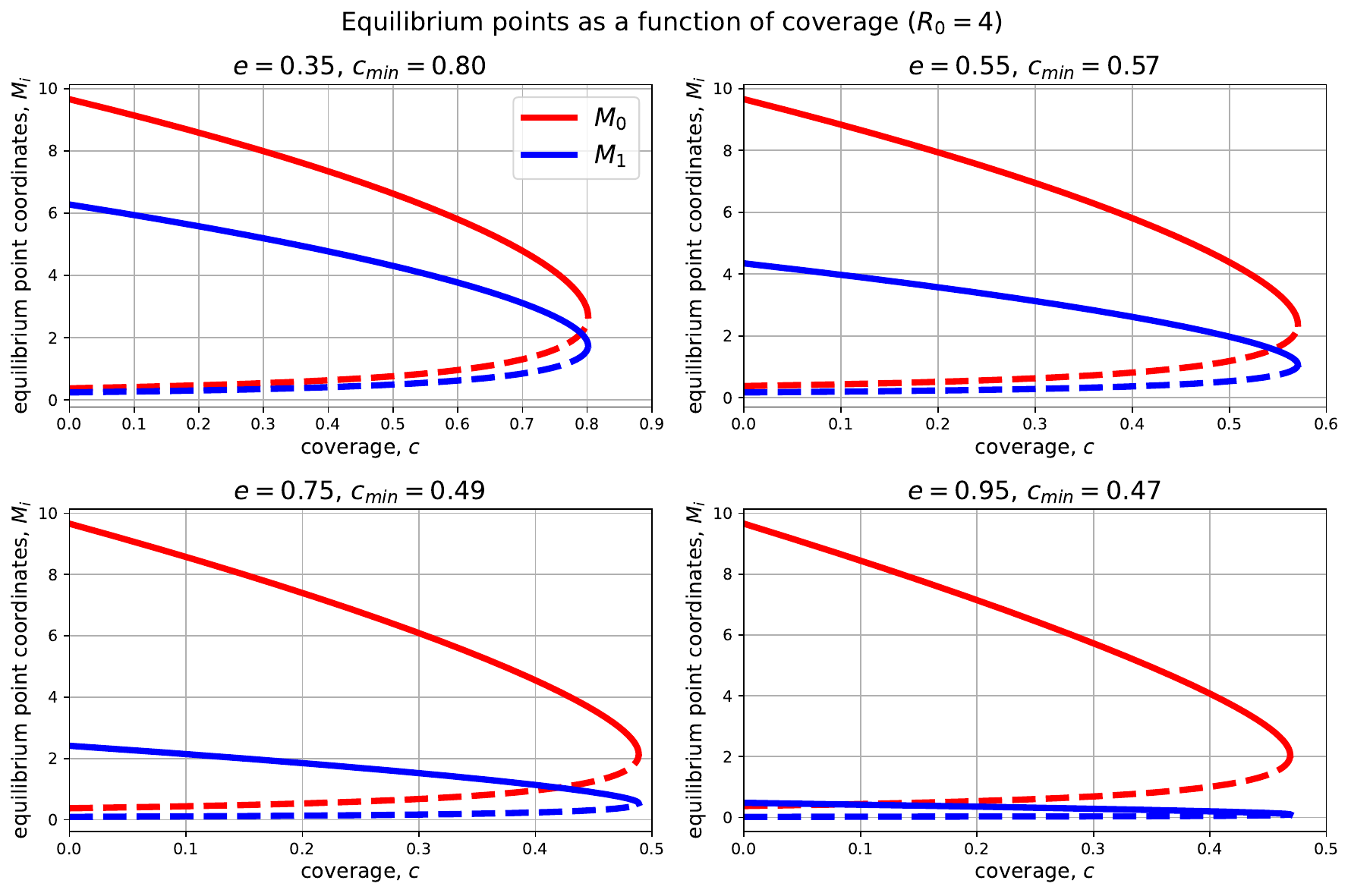}  
\caption{This bifurcation diagram shows the equilibria of the system as a function of coverage (\(c\)). The solid curves represent stable branches, and the dashed curves represent unstable branches. The analysis considers different values of effectiveness ($e=0.35$, 0.55, 0.75, and 0.95), with the following fixed parameters \(\alpha=0.574\), \(k=0.7\), \(z=0.92\), and \(R_0=4\) \citep{anderson1992infectious,seo1979egg}. 
The diagram shows the minimum coverage ($c_{\min}$) needed to reach a disease-free equilibrium at different levels of intervention effectiveness.}  
\label{f:coverage}  
\end{figure}  

In this case, we plot the bifurcation diagram as a function of coverage (\(c\)), while keeping the other parameters constant, as shown in Figure \ref{f:coverage}.  

Each curve represents the equilibrium values for a specific level of effectiveness.
The solid curves correspond to the stable branch of equilibria and the dashed curves correspond to the unstable branch.  

The equilibria of system \eqref{model2} are displayed as a function of \(c\) for different fixed values of effectiveness: \(e=0.35, 0.55, 0.75, 0.95\), and \(R_0 = 4\). 

This analysis reveals the minimum coverage thresholds required to drive the system toward a disease-free equilibrium, depending on the intervention effectiveness.

For example, when the intervention's effectiveness is set at \(35\%\) ($e=0.35$), achieving a disease-free equilibrium requires more than \(80\%\) ($c=0.80$) coverage of the population.  

As the effectiveness increases to 55\%, the required coverage decreases significantly, falling to 57\%.  

Further increases in effectiveness lead to even greater reductions. For effectiveness levels of 75\% and 95\%, the required coverage falls to approximately 49\% and 47\%, respectively.  

These results suggest that 75\% effectiveness is nearly equivalent to 95\% effectiveness in terms of the coverage needed to achieve disease-free equilibrium in a community.

Notably, if the WASH program's coverage remains below 47\%, it is unlikely that a disease-free equilibrium will be achieved, assuming that the initial transmission conditions ($R_0 = 4$) remain unchanged.

These findings emphasize the importance of ensuring a sufficiently high intervention effectiveness to reduce the required coverage and, consequently, effectively control and eliminate infection.

\subsubsection{Effectiveness}  

\begin{figure}[h!]  
\centering  
\includegraphics[width=0.7\linewidth]{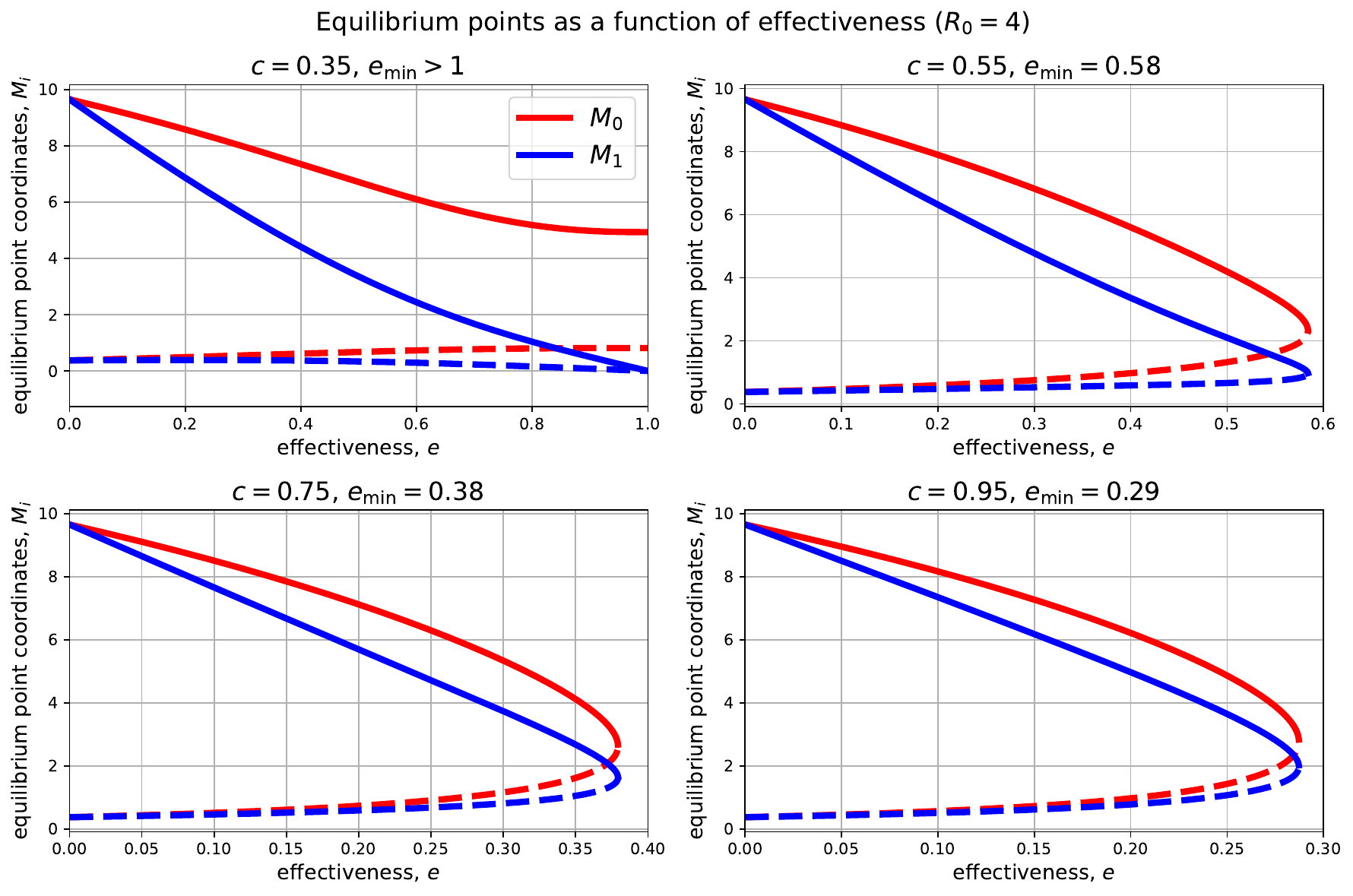}
\caption{This bifurcation diagram shows the equilibria of the system as a function of intervention effectiveness (\(e\)). The solid curves represent stable branches, and the dashed curves represent unstable branches. The analysis considers different values of coverage ($c=0.25$, 0.5, 0.75, and 0.95), with the following fixed parameters \(\alpha=0.574\), \(k=0.7\), \(z=0.92\), and \(R_0=4\) \citep{anderson1992infectious,seo1979egg}. The diagram shows the minimum effectiveness ($e_{\min}$) required to reach a disease-free equilibrium at different levels of intervention coverage.}  
\label{f:effectiveness}  
\end{figure}  

In this section, we analyze the bifurcation diagram as a function of intervention effectiveness (\(e\)), while keeping the coverage (\(c\)) fixed at various levels.  

Figure \ref{f:effectiveness} illustrates the system's equilibria as effectiveness \(e\) varies, considering different fixed coverage values ($c$) and maintaining \(R_0 = 4\).  

For a coverage level of $35\%$ ($c = 0.35$), the system is unlikely to reach a disease-free equilibrium unless external perturbations (not generated by WASH programs) push the state of the system below a critical threshold, or breakpoint.

However, as coverage increases, the effectiveness required  for elimination decreases. 
When coverage is set at 55\%, at least 58\% effectiveness is needed to eliminate the infection. 
Increasing coverage to 75\% reduces this threshold to 38\%, and at 95\% coverage, only 29\% effectiveness is needed to achieve a disease-free equilibrium.

These results highlight an inverse relationship between coverage and effectiveness: higher coverage requires less effectiveness required for elimination. This emphasizes the importance of maximizing both coverage and effectiveness in WASH programs to increase infection control and achieve eradication more efficiently.    


\subsubsection{Basic Reproduction Number}
\begin{figure}[h!]
\centering
\includegraphics[width=0.7\linewidth]{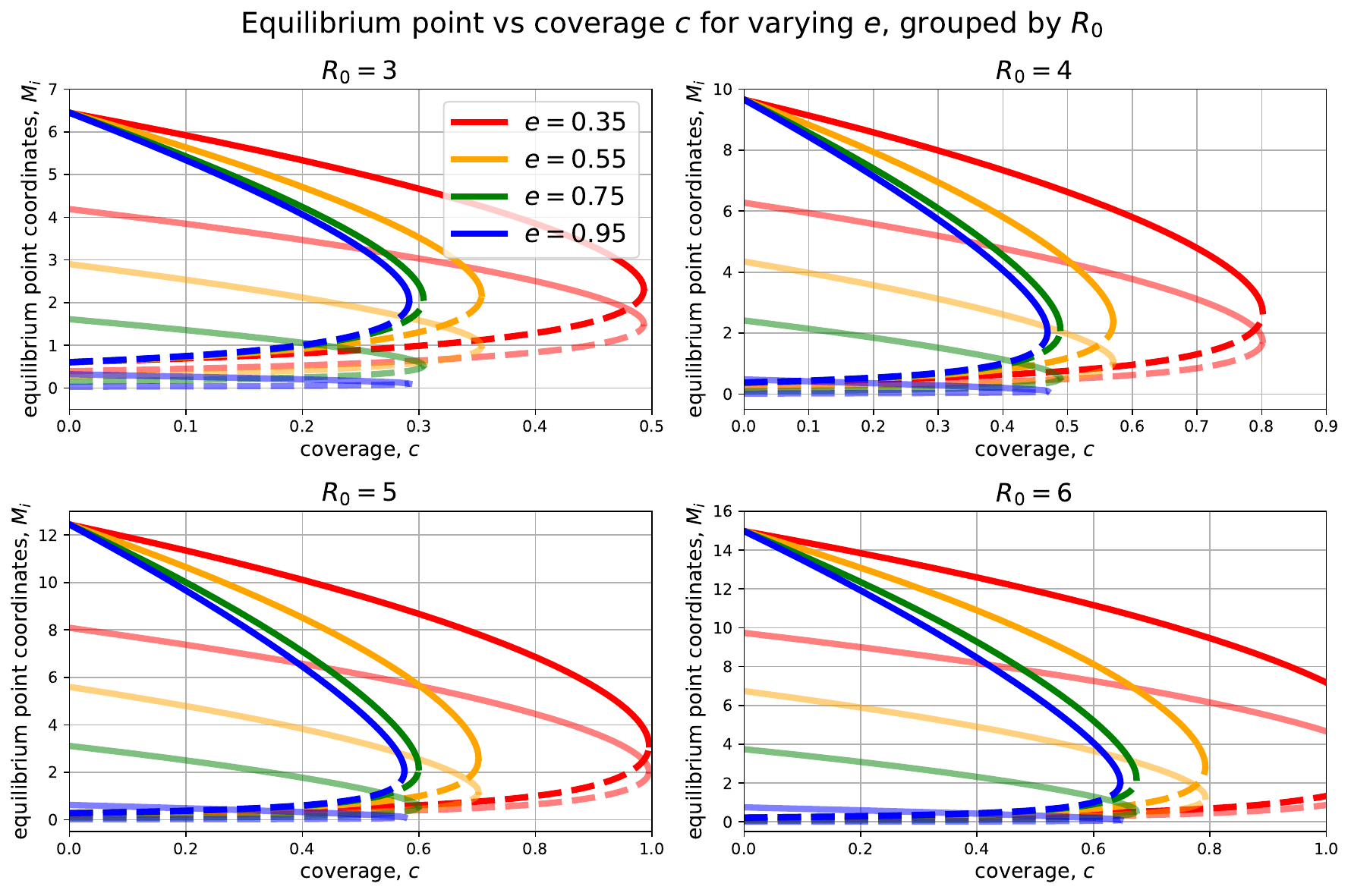}
\caption{
Bifurcation diagrams illustrate the equilibrium dynamics in relation to the intervention coverage ($c$). 
Each panel corresponds to a different value of the basic reproduction number ($R_0 = 3$, $4$, $5$, or $6$), 
with fixed intervention effectiveness levels ($e = 0.35$, $0.55$, $0.75$, or $0.95$).
The colored curves represent the equilibrium coordinates ($M_i$) of the dynamical system at each effectiveness level ($e$).
For each color, the solid curve corresponds to the untreated population ($M_0$), and the lighter curve corresponds to the treated population ($M_1$).
Continuous curves indicate stable equilibria, and dashed curves indicate unstable branches (breakpoints).
}
\label{f:plots_bifur_R0(coverage)}
\end{figure}

\begin{figure}
\centering
\includegraphics[width=0.7\linewidth]{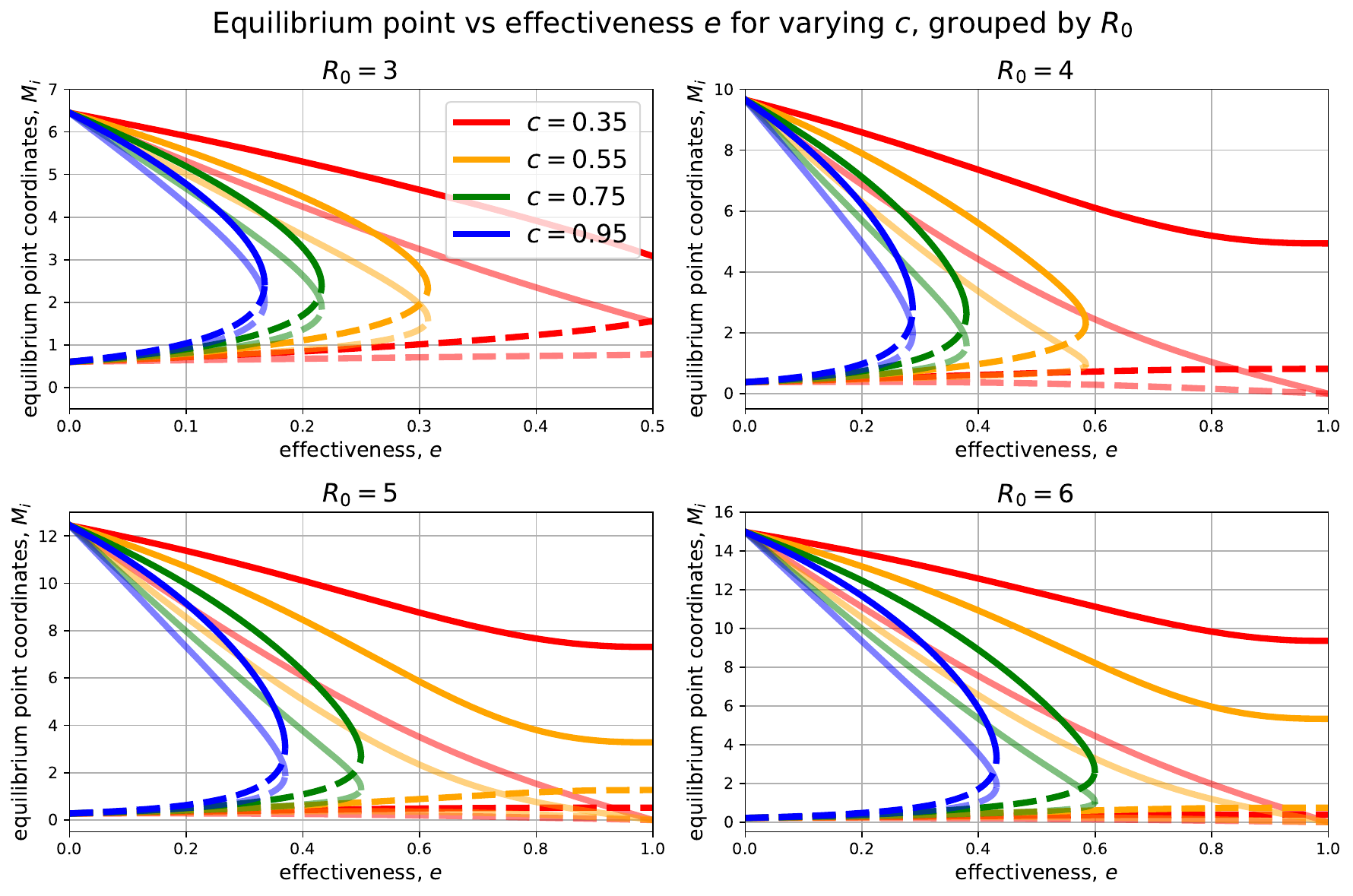}
\caption{
Bifurcation diagrams illustrate the equilibrium dynamics in relation to the intervention effectiveness ($e$). 
Each panel corresponds to a different value of the basic reproduction number ($R_0 = 3$, $4$, $5$, or $6$), 
with fixed intervention coverage levels ($c = 0.35$, $0.55$, $0.75$, or $0.95$).
The colored curves represent the equilibrium coordinates ($M_i$) of the dynamical system at each effectiveness level ($e$).
For each color, the solid curve corresponds to the untreated population ($M_0$), and the lighter curve corresponds to the treated population ($M_1$).
Continuous curves indicate stable equilibria, and dashed curves indicate unstable branches (breakpoints).
}
\label{f:plots_bifur_R0(effectiveness)}
\end{figure}

In the final scenario, we focus on the impact of the basic reproduction number ($R_0$), which reflects the community’s underlying transmission conditions prior to any intervention. In other words, $R_0$ represents the parasite’s transmission potential under endemic conditions, in the absence of control strategies. We analyze how different values of $R_0$ affect the ability of WASH interventions to eliminate the infection, considering fixed levels of program coverage ($c$) and effectiveness ($e$).

Figure~\ref{f:plots_bifur_R0(coverage)}  illustrates how the mean parasite burdens at equilibrium change with WASH program coverage at different basic reproduction numbers ($R_0$). The results indicate that higher $R_0$ values require significantly greater intervention coverage to achieve elimination. 

Specifically, infection elimination becomes unattainable at low intervention effectiveness levels ($e = 0.35$, for example) for $R_0 = 5$ or $6$ unless additional external interventions push the system below the transmission breakpoint.

Figure~\ref{f:plots_bifur_R0(effectiveness)} illustrates how changes in intervention effectiveness $e$ affect the equilibrium dynamics at different values of the basic reproduction number $R_0$. The diagrams reveal that higher $R_0$ values require substantially higher effectiveness to eliminate the infection. For example, when $R_0 = 5$ or $6$, elimination is not possible for any value of $e$ if the coverage remains below moderate levels (e.g., $c = 0.55$). This highlights the importance of achieving both high coverage and high effectiveness when facing highly transmissible scenarios.

These results highlight the central role of the basic reproduction number $R_0$ in determining the feasibility of helminth infection elimination through WASH interventions. 

As $R_0$ increases, progressively higher levels of program coverage and effectiveness are required to reach the bifurcation threshold beyond which endemic equilibria collapse. 

For moderate $R_0$ values (e.g., between 3 and 4), WASH programs with intermediate coverage and effectiveness can achieve elimination. 

However, in settings with high transmission ($R_0\ge5$), even with near-optimal implementation, WASH alone may be insufficient, thus requiring complementary strategies, such as mass drug administration (MDA).

This highlights the importance of tailoring control strategies according to underlying transmission intensity and considering integrated interventions to achieve the WHO's 2030 elimination  goals \citep{who2019sth2030,who2021roadmap}.


\subsubsection{Parasite aggregation parameter}
\label{sss:k}
\begin{figure}[h]
    \centering
    \begin{subfigure}{0.5\linewidth}
        \centering
        \includegraphics[width=\linewidth]{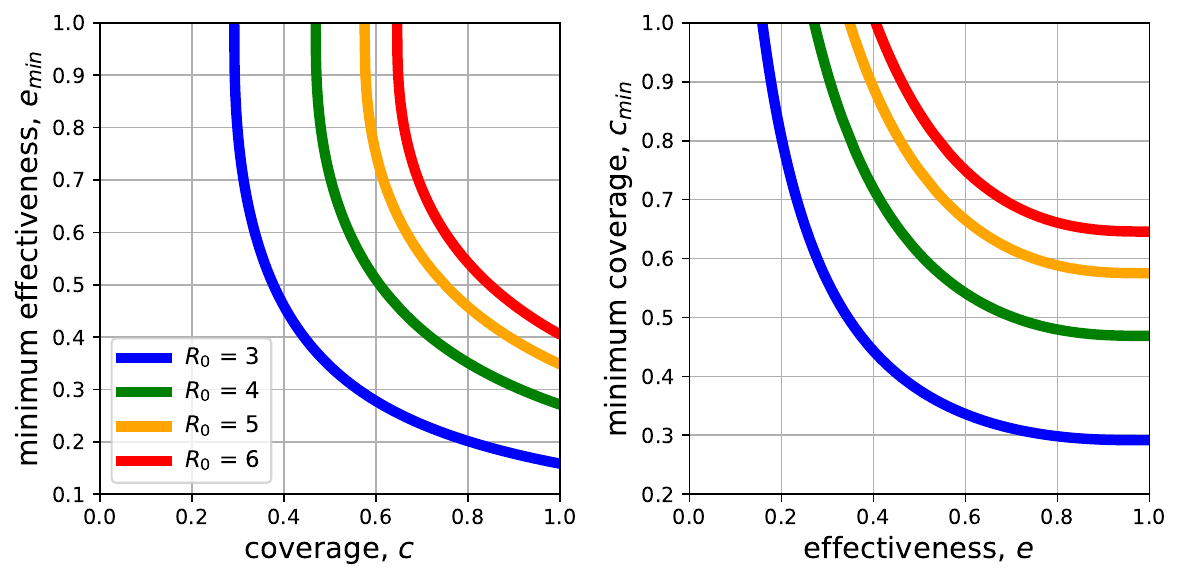}
    \end{subfigure}
    \vspace{0.1cm}  
    \begin{subfigure}{0.5\linewidth}
        \centering
        \includegraphics[width=\linewidth]{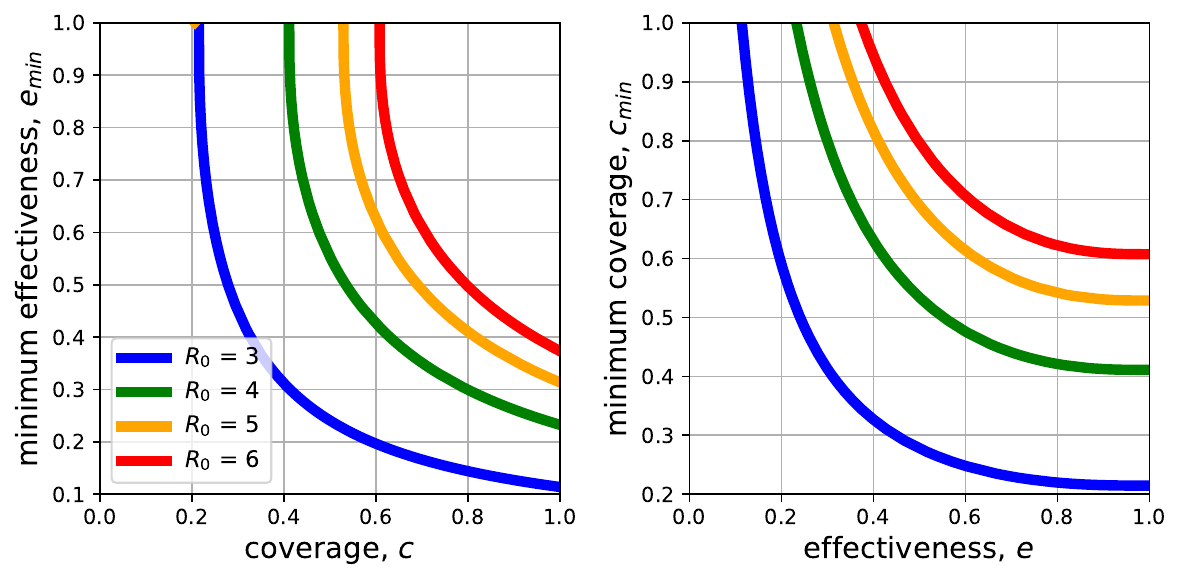}
    \end{subfigure}
    \caption{
    The left panel shows the minimum effectiveness ($e_{\min}$) for different values of coverage ($c$) assumed for the host population.  
    The right panel shows the minimum coverage ($c_{\min}$) for different values of effectiveness ($e$).  
    Each curve corresponds to a different value of $R_0 = 3$, 4, 5, or 6.  
    The upper panels correspond to a low aggregation value ($k = 0.07$), and the lower panels correspond to a high value ($k = 0.7$)~\citep{anderson1992infectious}.}
    \label{f:panel_completo}
\end{figure}

Figure~\ref{f:panel_completo} presents how intervention coverage and effectiveness interact to achieve control of helminth transmission under varying basic reproduction numbers ($R_0$) and parasite aggregation levels ($k$) in the host community.

The left panels illustrate the minimum effectiveness ($e_{\min}$) required to achieve control at different coverage levels ($c$). 
As $R_0$ increases, the required effectiveness also increases, reflecting the greater challenge of controlling highly transmissible infections. 

The parasite aggregation parameter $k$ also plays a critical role. When aggregation is low ($k = 0.07$), achieving control requires higher effectiveness than when aggregation is high ($k = 0.7$), for comparable coverage levels. This suggests that higher aggregation facilitates control by concentrating infections among fewer hosts.

On the other hand, the right panels show the minimum coverage required ($c_{\min}$) to achieve control at different levels of intervention effectiveness. As in the left panels, higher values of $R_0$ require greater coverage to compensate for reduced effectiveness. When parasite aggregation is high, the required coverage is lower, especially for infections with high transmission potential.

These results highlight the value of taking into account parasite aggregation when designing control strategies. For infections with high $R_0$, both high coverage and effectiveness are essential. However, high aggregation can ease these requirements by focusing control efforts on the most heavily infected individuals. This insight can inform more efficient and context-sensitive public health interventions.

\section{Numerical Simulations}

\begin{figure}[h]
\centering     \includegraphics[width=0.7\linewidth]{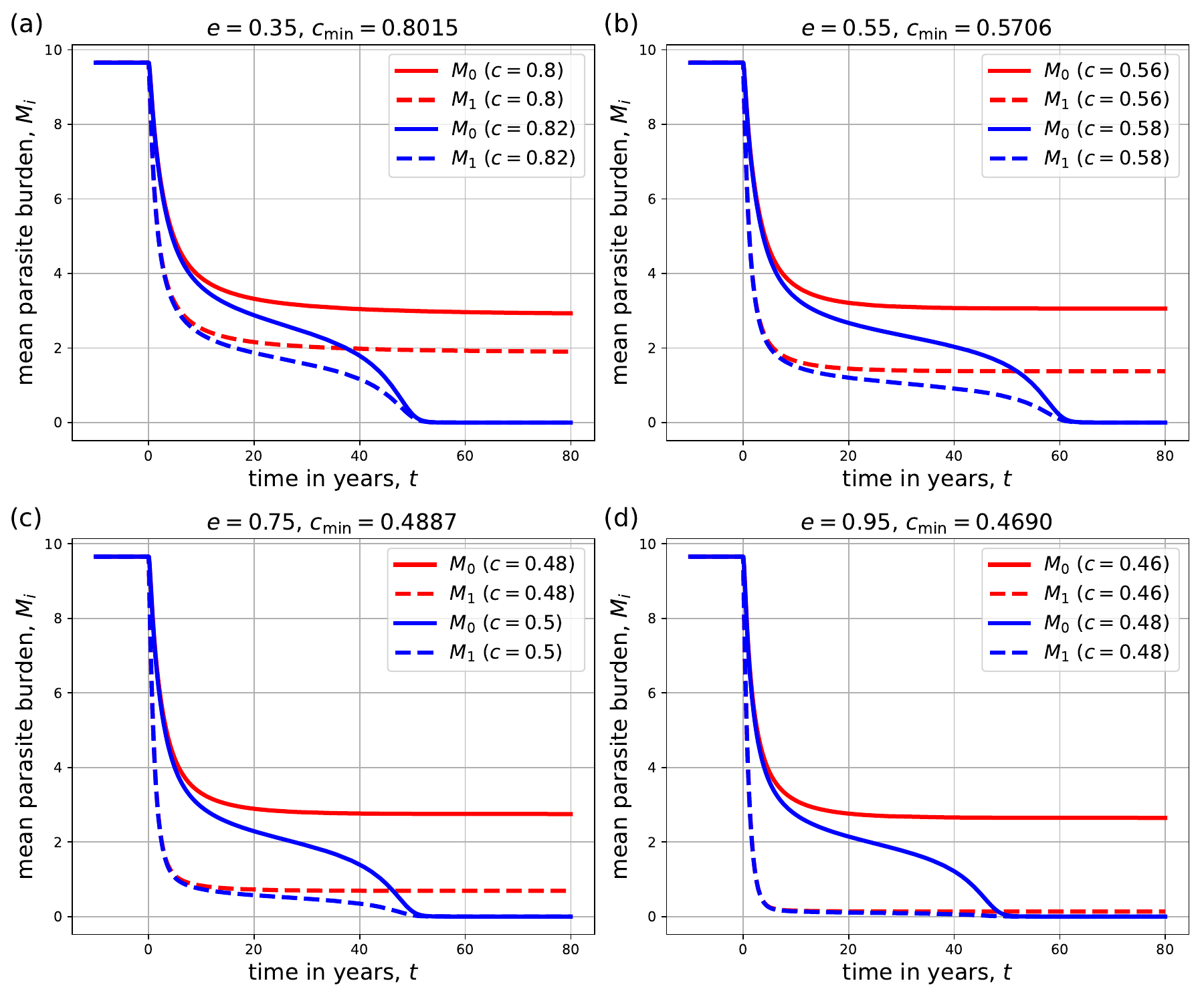}
\caption{Infection dynamics under different WASH effectiveness levels ($e$ = 0.35, 0.55, 0.75, and 0.95), shown in panels \textsf{(a)}--\textsf{(d)}, respectively. For each effectiveness level, the minimum required coverage threshold $c_{\min}$ is estimated. In all cases, when $c > c_{\min}$, the system evolves toward disease-free equilibrium, indicating successful eradication. The simulations are conducted for $R_0 = 4$, assuming an initial endemic equilibrium. The parameters used were $\alpha = 0.574$, $k = 0.7$, and $z = 0.92$ \citep{anderson1992infectious,seo1979egg}.}
\label{f:simu1}
\end{figure}

\begin{figure}[h]
\centering
\includegraphics[width=0.7\linewidth]{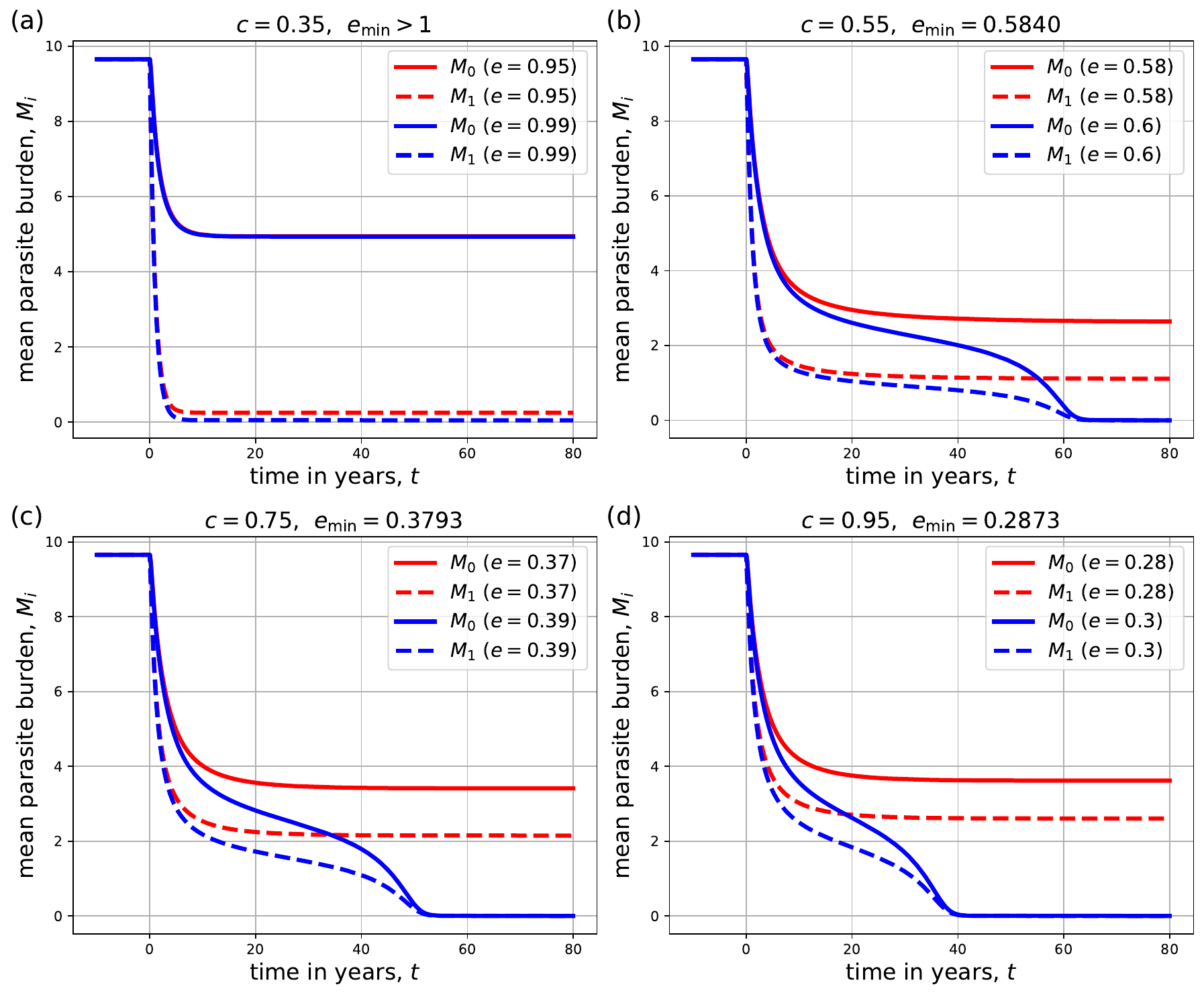}
\caption{Infection dynamics under different WASH coverage levels ($c = 0.35$, 0.55, 0.75, and 0.95), shown in panels \textsf{(a)}--\textsf{(d)}, respectively. For each coverage level, the minimum required effectiveness threshold $e_{\min}$ is estimated. In cases where $e < e_{\min}$, the infection persists, whereas for $e > e_{\min}$, the system evolves toward disease-free equilibrium. Notably, in panel \textsf{(a)} ($c = 0.35$), the value of $e_{\min}$ exceeds one, indicating that eradication is not achievable through WASH effectiveness alone at such low coverage. The simulations are conducted for $R_0 = 4$, assuming an initial endemic equilibrium. The parameters used were $\alpha = 0.574$, $k = 0.7$, and $z = 0.92$ \citep{anderson1992infectious,seo1979egg}.}
\label{f:simu2}
\end{figure}

In this section, we perform numerical simulations to study the infection dynamic behavior under the influence of WASH program interventions. 

We aim to evaluate the impact of different values of  WASH program parameters, such as coverage ($c$) and effectiveness ($e$), on the long-term behavior of the system, particularly in relation to the eradication or persistence of the infection in the host community.

Specifically, we aim to determine the parameter thresholds, denoted by $c_{\min}$ and $e_{\min}$, that differentiate conditions under which the infection persists from those that lead to eradication, as discussed in Section~\ref{s:model1}.


We simulated a scenario with a basic reproduction number $R_0 = 4$, assuming that the infection had already reached its endemic equilibrium, as described in Section~\ref{s:model1}.
This value of $R_0$ represents a moderately high transmission setting and allows us to explore realistic control scenarios. Additionally, we maintain the parasite aggregation parameter at $k = 0.7$, which is within the typical biological range and is consistent with previous studies \citep{anderson1992infectious}.

Although Section~\ref{s:impact} presents a broader parametric analysis for different values of $R_0$, $c$, $e$, and $k$, we focus here on a representative case to illustrate the key dynamical patterns. The results and thresholds obtained in this section can be generalized to other settings with similar qualitative behavior. Our goal is to demonstrate that exceeding the minimum intervention thresholds is sufficient to drive the system toward disease-free equilibrium.

At time $t = 0$, the WASH program intervention is implemented in the host community, modifying the infection dynamics as described by the extended model in Section~\ref{s:model2}. 

The simulations explore the impact of different combinations of coverage ($c$) and effectiveness ($e$) on the long-term dynamics of infection within the host community. We considered values of 35\%, 55\%, 75\%, and 95\% for each parameter, as illustrated in Figures ~\ref{f:simu1} and ~\ref{f:simu2}, respectively.


Figure~\ref{f:simu1} shows the model dynamics for two values of $c$ near the threshold $c_{\min}$. 
For values of $c < c_{\min}$, the system stabilizes at an endemic equilibrium, indicating that the infection persists in the host community. 
In contrast, for $c > c_{\min}$, the infection tends toward disease-free equilibrium, which leads to disease eradication.

Similarly, we analyze the system behavior around the effectiveness threshold $e_{\min}$. As with $c_{\min}$, the system exhibits threshold behavior: When the effectiveness $e$ is below this critical value, the infection persists, and the system stabilizes at endemic equilibrium. In contrast, for $e > e_{\min}$, the system evolves toward a disease-free equilibrium, indicating successful eradication (see Figure~\ref{f:simu2}).

In Figure~\ref{f:simu2} \textsf{(a)}, for a fixed coverage level of $c = 0.35$, we simulate unrealistically high levels of effectiveness. However, we did not observe a transition to disease-free equilibrium. This outcome occurs because, in this case, $e_{\min} $ is greater than one.
This means that even perfect WASH program effectiveness would not be sufficient to eliminate the infection. 

These findings highlight that although WASH programs can alter system dynamics, they may not be sufficient to eliminate infection unless the necessary thresholds are achievable. In such cases, external interventions are necessary.

Figures ~\ref{f:simu1} and ~\ref{f:simu2} show that eradication may still take a considerable amount of time, even when $c > c_{\min}$ and $e > e_{\min}$, for different levels of WASH coverage and effectiveness (35\%, 55\%, 75\%, and 95\%). These prolonged transients reflect the system’s slow convergence toward a disease-free state and highlight the limitations of relying solely on behavioral interventions.

While the model does not provide quantitative predictions regarding the time required for eradication, it suggests that incorporating external measures, such as antiparasitic treatments, can accelerate the process. These treatments reduce the mean parasite burden and push the system past the bifurcation threshold into the basin of attraction of the disease-free equilibrium more efficiently.

In summary, the results emphasize the importance of combining control strategies, such as WASH programs and MDA (Mass Drug Administration) programs, to achieve effective and timely disease control.


\section{Discussion and Conclusions}

This study presents a novel mathematical framework that isolates and quantifies the impact of community-based water, sanitation, and hygiene (WASH) program interventions on the transmission dynamics of soil-transmitted helminth (STH) infections. Unlike previous studies that analyzed these programs in conjunction with mass drug administration (MDA) \citep{coffeng2018predicted}, our approach allows us to evaluate the independent contribution of WASH interventions to helminth infection control \citep{who2019sth2030,who2021roadmap}.

Our findings highlight that both the coverage and effectiveness of WASH programs are critical, as high values in one cannot compensate for very low values in the other. These results are consistent with empirical observations \citep{freeman2017impact,garn2022interventions,ziegelbauer2012effect}.

We derived analytical expressions for the basic and effective reproduction numbers ($R_0$ and $R_e$) by explicitly incorporating the intervention coverage ($c$) and effectiveness ($e$) as model parameters. We also performed a bifurcation analysis to determine the thresholds required to eliminate infections.

The model confirms that a saddle-node bifurcation governs the transition between endemic and disease-free equilibria, as previously described in classical macroparasite models \citep{anderson1985helminth,anderson1992infectious,chan1994development}. 
These results demonstrate the presence of multiple equilibria: a stable endemic equilibrium ($E^s$), a stable disease-free equilibrium ($E^0$), and an unstable equilibrium ($E^u$) that separates them. This insight is significant because it shows that the basic reproduction number alone does not fully describe infection persistence. Rather, persistence depends on the initial conditions relative to a critical transmission breakpoint \citep{diekmann1990definition}.

Our findings show that achieving a disease-free equilibrium ($E^0$) requires surpassing the minimum coverage ($c_{\min}$) and effectiveness ($e_{\min}$) thresholds. These thresholds are not constant but vary as nonlinear functions of the model parameters and initial transmission intensity ($R_0$). Furthermore, the model reveals a compensatory relationship between coverage and effectiveness: high effectiveness can compensate for low coverage, and vice versa.

From a policy perspective, the bifurcation diagrams generated in Section~\ref{s:impact} provide practical tools to support decision-making. In particular, the curves shown in Section~\ref{sss:k} (Figure~\ref{f:panel_completo}) summarize the minimum effectiveness ($e_{\min}$) and coverage ($c_{\min}$) needed to eliminate infection under different values of the basic reproduction number ($R_0$) and parasite aggregation ($k$). 
This figure is an important part of the study because it provides a concise visualization of feasible control targets that public health officials can adapt to specific epidemiological settings. For example, in highly aggregated populations (high $k$), lower coverage may still suffice, offering strategic leverage in settings with limited resources.

In addition, the model aligns with the WHO's strategic objectives of eliminating STH as a public health problem by 2030 \citep{who2019sth2030,who2021roadmap}. However, empirical observations suggest that relying solely on mass drug administration (MDA),  particularly in high-transmission settings, may be insufficient. Instead, integrated approaches combining MDA with sustained and effective WASH interventions are necessary, particularly in communities with high $R_0$ values. 
These results are consistent with those of large-scale cluster trials and community studies (\citep{nery2019wash, clarke2018s, hurlimann2018effect}), highlighting the importance of combining pharmacological and structural interventions.

Methodologically, this work highlights the value of bifurcation analysis as a tool for guiding public health interventions. Unlike traditional threshold-based models, our framework captures the nonlinear and sometimes counterintuitive behavior of macroparasite transmission systems, particularly in heterogeneous populations with partially implemented interventions.

Although studies such as those by \citep{coffeng2018predicted} suggest that the short-term impact of WASH alone is limited, our results show that, under the right conditions of coverage ($c$) and effectiveness ($e$), WASH interventions can drive the system toward elimination.
In scenarios where WASH is insufficient, complementary strategies such as MDA are necessary to temporarily reduce the parasite burden and shift the system below the transmission breakpoint, where WASH can maintain elimination \citep{coffeng2018predicted}. 


These results are consistent with empirical evidence showing the limited short-term impact of WASH when implemented in isolation in high-transmission environments \citep{freeman2017impact, ziegelbauer2012effect}, while also supporting findings that emphasize its importance for long-term control and sustainability when community coverage is high.

Nevertheless, the model has limitations. It assumes homogeneous behavior within intervention and non-intervention subpopulations, and it does not account for stochastic variability or adaptive feedback in host community behavior. In addition, the model lacks spatial structure and does not account for migration or focal transmission. Future extensions could incorporate the stochasticity and impulsive dynamics associated with periodic MDA, as well as spatial and age heterogeneity, using metapopulation frameworks or partial differential equations.

Importantly, the application of these findings requires context-specific calibration. Social, economic, and infrastructural conditions can significantly impact the achievable coverage and real-world effectiveness of WASH interventions. Therefore, while the thresholds identified here provide general guidance, they must be interpreted in light of local epidemiological data and operational constraints.

In conclusion, this modeling framework improves our theoretical understanding of helminth transmission dynamics under WASH interventions and provides actionable implementation insights.

Future work should explore model extensions that incorporate behavioral feedback, economic costs, and interactions with other control strategies.

Structural interventions, such as WASH programs, can achieve and maintain elimination when implemented and sustained appropriately, even in the absence of pharmacological measures.

By combining this framework with periodic MDA and spatial models, we provide a powerful tool for integrated disease management that aligns with the World Health Organization's 2030 goals \cite{who2019sth2030,who2021roadmap}.


\section*{Aknowledgements}
This research was partially supported by CIUNSA grants 2024-2982 and 2024-2883. JPA is a member of CONICET, and GML is a CONICET postdoctoral fellow.

\section*{Conflict of interest}
The authors declare no conflicts of interest.


\renewcommand{\bibname}{References}
\bibliographystyle{apalike}
\bibliography{references}

\begin{thebibliography}{}

\bibitem[Anderson and May, 1985]{anderson1985helminth}
Anderson, R.~M. and May, R.~M. (1985).
\newblock Helminth infections of humans: mathematical models, population
  dynamics, and control.
\newblock {\em Advances in parasitology}, 24:1--101.

\bibitem[Anderson and May, 1992]{anderson1992infectious}
Anderson, R.~M. and May, R.~M. (1992).
\newblock {\em Infectious diseases of humans: dynamics and control}.
\newblock Oxford University Press.

\bibitem[Basanez et~al., 2012]{basanez2012research}
Basanez, M.~G., McCarthy, J.~S., French, M.~D., Yang, G.~J., Walker, M.,
  Gambhir, M., Prichard, R.~K., and Churcher, T.~S. (2012).
\newblock A research agenda for helminth diseases of humans: modelling for
  control and elimination.
\newblock {\em PLoS Neglected Tropical Diseases}, 6(4):e1548.

\bibitem[Chan et~al., 1994]{chan1994development}
Chan, M., Guyatt, H., Bundy, D.~A., and Medley, G. (1994).
\newblock The development and validation of an age-structured model for the
  evaluation of disease control strategies for intestinal helminths.
\newblock {\em Parasitology}, 109(3):389--396.

\bibitem[Chicone, 2006]{chicone2006ordinary}
Chicone, C.~C. (2006).
\newblock {\em Ordinary differential equations with applications}, volume~34.
\newblock Springer.

\bibitem[Churcher et~al., 2006]{churcher2006density}
Churcher, T.~S., Filipe, J.~A., and Bas{\'a}{\~n}ez, M.~G. (2006).
\newblock Density dependence and the control of helminth parasites.
\newblock {\em Journal of animal ecology}, pages 1313--1320.

\bibitem[Clarke et~al., 2018]{clarke2018s}
Clarke, N.~E., Clements, A.~C., Amaral, S., Richardson, A., McCarthy, J.~S.,
  McGown, J., Bryan, S., Gray, D.~J., and Nery, S.~V. (2018).
\newblock {(S)WASH-D} for worms: A pilot study investigating the differential
  impact of school-versus community-based integrated control programs for
  soil-transmitted helminths.
\newblock {\em PLOS Neglected Tropical Diseases}, 12(5):e0006389.

\bibitem[Clasen et~al., 2014]{clasen2014effectiveness}
Clasen, T., Boisson, S., Routray, P., Torondel, B., Bell, M., Cumming, O.,
  Ensink, J., Freeman, M., Jenkins, M., Odagiri, M., et~al. (2014).
\newblock Effectiveness of a rural sanitation programme on diarrhoea,
  soil-transmitted helminth infection, and child malnutrition in odisha, india:
  a cluster-randomised trial.
\newblock {\em The Lancet Global Health}, 2(11):e645--e653.

\bibitem[Coffeng et~al., 2018]{coffeng2018predicted}
Coffeng, L., Vaz~Nery, S., Gray, D., Bakker, R., de~Vlas, S., and Clements, A.
  (2018).
\newblock Predicted short and long-term impact of deworming and water, hygiene,
  and sanitation on transmission of soil-transmitted helminths.
\newblock {\em PLoS Neglected Tropical Diseases}, 12(12):e0006758.

\bibitem[Diekmann and Heesterbeek, 2000]{diekmann2000mathematical}
Diekmann, O. and Heesterbeek, J. A.~P. (2000).
\newblock {\em Mathematical epidemiology of infectious diseases: model
  building, analysis and interpretation}, volume~5.
\newblock John Wiley \& Sons.

\bibitem[Diekmann et~al., 1990]{diekmann1990definition}
Diekmann, O., Heesterbeek, J. A.~P., and Metz, J. A.~J. (1990).
\newblock On the definition and the computation of the basic reproduction ratio
  {$R_0$} in models for infectious diseases in heterogeneous populations.
\newblock {\em Journal of mathematical biology}, 28:365--382.

\bibitem[Freeman et~al., 2017]{freeman2017impact}
Freeman, M.~C., Garn, J.~V., Sclar, G.~D., Boisson, S., Medlicott, K.,
  Alexander, K.~T., Penakalapati, G., Anderson, D., Mahtani, A.~G., Grimes,
  J.~E., et~al. (2017).
\newblock The impact of sanitation on infectious disease and nutritional
  status: a systematic review and meta-analysis.
\newblock {\em International journal of hygiene and environmental health},
  220(6):928--949.

\bibitem[Garn et~al., 2022]{garn2022interventions}
Garn, J.~V., Wilkers, J.~L., Meehan, A.~A., Pfadenhauer, L.~M., Burns, J.,
  Imtiaz, R., and Freeman, M.~C. (2022).
\newblock Interventions to improve water, sanitation, and hygiene for
  preventing soil-transmitted helminth infection.
\newblock {\em Cochrane Database of Systematic Reviews}, (6).

\bibitem[Hall and Holland, 2000]{hall2000geographical}
Hall, A. and Holland, C. (2000).
\newblock Geographical variation in ascaris lumbricoides fecundity and its
  implications for helminth control.
\newblock {\em Parasitology Today}, 16(12):540--544.

\bibitem[Heesterbeek and Roberts, 1995]{heesterbeek1995threshold}
Heesterbeek, J. A.~P. and Roberts, M.~G. (1995).
\newblock Threshold quantities for helminth infections.
\newblock {\em Journal of mathematical biology}, 33:415--434.

\bibitem[Hotez et~al., 2014]{hotez2014global}
Hotez, P.~J., Alvarado, M., Bas{\'a}{\~n}ez, M.-G., Bolliger, I., Bourne, R.,
  Boussinesq, M., Brooker, S.~J., Brown, A.~S., Buckle, G., Budke, C.~M.,
  et~al. (2014).
\newblock The global burden of disease study 2010: interpretation and
  implications for the neglected tropical diseases.
\newblock {\em PLoS neglected tropical diseases}, 8(7):e2865.

\bibitem[H{\"u}rlimann et~al., 2018]{hurlimann2018effect}
H{\"u}rlimann, E., Silu{\'e}, K.~D., Zouzou, F., Ouattara, M., Schmidlin, T.,
  Yapi, R.~B., Houngbedji, C.~A., Dongo, K., Kouadio, B.~A., Kon{\'e}, S.,
  et~al. (2018).
\newblock Effect of an integrated intervention package of preventive
  chemotherapy, community-led total sanitation and health education on the
  prevalence of helminth and intestinal protozoa infections in c{\^o}te
  d’ivoire.
\newblock {\em Parasites \& vectors}, 11:1--20.

\bibitem[Landeryou et~al., 2022]{landeryou2022longitudinal}
Landeryou, T., Maddren, R., Rayment~Gomez, S., Kalahasti, S., Liyew, E.~F.,
  Chernet, M., Mohammed, H., Wuletaw, Y., Truscott, J., Phillips, A.~E., et~al.
  (2022).
\newblock Longitudinal monitoring of prevalence and intensity of
  soil-transmitted helminth infections as part of community-wide mass drug
  administration within the geshiyaro project in the bolosso sore district,
  wolaita, ethiopia.
\newblock {\em PLoS Neglected Tropical Diseases}, 16(9):e0010408.

\bibitem[Lopez and Aparicio, 2022]{lopez2022modeling}
Lopez, G.~M. and Aparicio, J.~P. (2022).
\newblock Modeling macroparasite infection dynamics.
\newblock {\em Journal of Mathematical Modeling of Biological Systems}.

\bibitem[Lopez and Aparicio, 2024]{lopez2024modeling}
Lopez, G.~M. and Aparicio, J.~P. (2024).
\newblock Mathematical modeling of mating probability and fertile egg
  production in helminth parasites.
\newblock {\em Bulletin of Mathematical Biology}.

\bibitem[Nery et~al., 2019]{nery2019wash}
Nery, S.~V., Traub, R.~J., McCarthy, J.~S., Clarke, N.~E., Amaral, S.,
  Llewellyn, S., Weking, E., Richardson, A., Campbell, S.~J., Gray, D.~J.,
  et~al. (2019).
\newblock Wash for worms: a cluster-randomized controlled trial of the impact
  of a community integrated water, sanitation, and hygiene and deworming
  intervention on soil-transmitted helminth infections.
\newblock {\em The American journal of tropical medicine and hygiene},
  100(3):750.

\bibitem[{Pan American Health Organization}, 2003]{paho2003}
{Pan American Health Organization} (2003).
\newblock Framework for a regional program for control of soil-transmitted
  helminth infections and schistosomiasis in the americas.
\newblock Technical report, Pan American Health Organization, Santo Domingo,
  Dominican Republic.

\bibitem[Patil et~al., 2014]{patil2014effect}
Patil, S.~R., Arnold, B.~F., Salvatore, A.~L., Briceno, B., Ganguly, S.,
  Colford~Jr, J.~M., and Gertler, P.~J. (2014).
\newblock The effect of india's total sanitation campaign on defecation
  behaviors and child health in rural madhya pradesh: a cluster randomized
  controlled trial.
\newblock {\em PLoS medicine}, 11(8):e1001709.

\bibitem[Perko, 2013]{perko2013differential}
Perko, L. (2013).
\newblock {\em Differential equations and dynamical systems}, volume~7.
\newblock Springer Science \& Business Media.

\bibitem[Puspita et~al., 2020]{puspita2020health}
Puspita, W.~L., Khayan, K., Hariyadi, D., Anwar, T., Wardoyo, S., and Ihsan,
  B.~M. (2020).
\newblock Health education to reduce helminthiasis: Deficits in diets in
  children and achievement of students of elementary schools at pontianak, west
  kalimantan.
\newblock {\em Journal of Parasitology Research}, 2020(1):4846102.

\bibitem[Seo et~al., 1979]{seo1979egg}
Seo, B.~S., Cho, S.~Y., and Chai, J.~Y. (1979).
\newblock Egg discharging patterns of ascaris lumbricoides in low worm burden
  cases.
\newblock {\em The Korean Journal of Parasitology}, 17(2):98--104.

\bibitem[Sotomayor, 1973]{sotomayor1973generic}
Sotomayor, J. (1973).
\newblock Generic bifurcations of dynamical systems.
\newblock In {\em Dynamical systems}, pages 561--582. Elsevier.

\bibitem[Steinmann et~al., 2015]{steinmann2015control}
Steinmann, P., Yap, P., Utzinger, J., Du, Z.-W., Jiang, J.-Y., Chen, R., Wu,
  F.-W., Chen, J.-X., Zhou, H., and Zhou, X.-N. (2015).
\newblock Control of soil-transmitted helminthiasis in yunnan province,
  people's republic of china: experiences and lessons from a 5-year
  multi-intervention trial.
\newblock {\em Acta tropica}, 141:271--280.

\bibitem[Strogatz, 2024]{strogatz2024nonlinear}
Strogatz, S.~H. (2024).
\newblock {\em Nonlinear dynamics and chaos: with applications to physics,
  biology, chemistry, and engineering}.
\newblock Chapman and Hall/CRC.

\bibitem[Strunz et~al., 2014]{strunz2014water}
Strunz, E.~C., Addiss, D.~G., Stocks, M.~E., Ogden, S., Utzinger, J., and
  Freeman, M.~C. (2014).
\newblock Water, sanitation, hygiene, and soil-transmitted helminth infection:
  a systematic review and meta-analysis.
\newblock {\em PLoS medicine}, 11(3):e1001620.

\bibitem[Ugwu et~al., 2024]{ugwu2024impact}
Ugwu, S.~C., Muoka, M.~O., MacLeod, C., Bick, S., Cumming, O., and Braun, L.
  (2024).
\newblock The impact of community based interventions for the prevention and
  control of soil-transmitted helminths: A systematic review and meta-analysis.
\newblock {\em PLOS Global Public Health}, 4(10):e0003717.

\bibitem[{World Health Organization}, 2011]{who2011helminth}
{World Health Organization} (2011).
\newblock Helminth control in school‑age children: A guide for managers of
  control programmes.
\newblock Technical report, World Health Organization, Geneva, Switzerland.

\bibitem[{World Health Organization}, 2012]{who2012soil}
{World Health Organization} (2012).
\newblock Soil-transmitted helminthiases: Eliminating soil-transmitted
  helminthiases as a public health problem in children: Progress report
  2001--2010 and strategic plan 2011--2020.
\newblock Technical report, World Health Organization, Geneva, Switzerland.

\bibitem[{World Health Organization}, 2019]{who2019sth2030}
{World Health Organization} (2019).
\newblock 2030 targets for soil-transmitted helminthiases control programmes.
\newblock Technical report, World Health Organization, Geneva, Switzerland.

\bibitem[{World Health Organization}, 2020]{who2021roadmap}
{World Health Organization} (2020).
\newblock Ending the neglect to attain the sustainable development goals: A
  road map for neglected tropical diseases 2021--2030.
\newblock Technical report, World Health Organization, Geneva, Switzerland.

\bibitem[Ziegelbauer et~al., 2012]{ziegelbauer2012effect}
Ziegelbauer, K., Speich, B., M{\"a}usezahl, D., Bos, R., Keiser, J., and
  Utzinger, J. (2012).
\newblock Effect of sanitation on soil-transmitted helminth infection:
  systematic review and meta-analysis.
\newblock {\em PLoS medicine}, 9(1):e1001162.

\end{thebibliography}
\end{document}